\documentclass[fleqn,usenatbib]{mnras}

\usepackage{newtxtext,newtxmath}

\usepackage[T1]{fontenc}

\DeclareRobustCommand{\VAN}[3]{#2}
\let\VANthebibliography\thebibliography
\def\thebibliography{\DeclareRobustCommand{\VAN}[3]{##3}\VANthebibliography}

\usepackage{graphicx}	
\usepackage{amsmath}	
\usepackage{makecell}
\usepackage{array}
\usepackage{subcaption}
\usepackage[export]{adjustbox}
\usepackage{xcolor}

\newcommand{\cfbox}[2]{\colorlet{currentcolor}{.}{\color{#1}\fbox{\color{currentcolor}#2}}}

\newcommand{\Mi}{ $M_{\text{ZAMS}}$ } 
\newcommand{\Z}{ $Z$ } 
\newcommand{\aOv}{ $\alpha_{\text{ov}}$ } 
\newcommand{\OmOmC}{ $\Omega / \Omega_{\text{crit}}$ } 
\newcommand{\aSC}{ $\alpha_{\text{sc}}$ }

\title{Predicting the Heaviest Black Holes below the Pair Instability Gap}

\author[E. R. J. Winch et al.]{
Ethan R. J. Winch,$^{1,2}$\thanks{E-mail: ethan.winch@armagh.ac.uk}
Jorick S. Vink,$^{1}$
Erin R. Higgins$^{1}$
and Gautham N. Sabhahit$^{1,2}$
\\
$^{1}$Armagh Observatory and Planetarium (AOP),
              Armagh, College Hill, BT61 9DB\\
$^{2}$School of Maths and Physics, Queen's University Belfast, Northern Ireland, University Road, BT7 1NN
}

\date{Accepted 2024 January 25. Received 2024 January 24; in original form 2023 June 14}

\pubyear{2024}

\begin{document}
\label{firstpage} 
\pagerange{\pageref{firstpage}--\pageref{lastpage}}
\maketitle

\begin{abstract}
Traditionally, the pair instability (PI) mass gap is located between 50\,and 130\,$M_{\odot}$, with stellar mass black holes (BHs) expected to "pile up" towards the lower PI edge. However, this lower PI boundary is based on the assumption that the star has already lost its hydrogen (H) envelope. With the announcement of an "impossibly" heavy BH of 85\,$M_{\odot}$ as part of GW\,190521 located inside the traditional PI gap, we realised that blue supergiant (BSG) progenitors with small cores but large Hydrogen envelopes at low metallicity ($Z$) could directly collapse to heavier BHs than had hitherto been assumed. The question of whether a single star can produce such a heavy BH is important, independent of gravitational wave events. Here, we systematically investigate the masses of stars inside the traditional PI gap by way of a grid of 336 detailed MESA stellar evolution models calculated across a wide parameter space, varying stellar mass, overshooting, rotation, semi-convection, and $Z$. We evolve low $Z$ stars in the range $10^{-3} < Z / Z_{\odot} < Z_{\rm SMC}$, making no prior assumption regarding the mass of an envelope, but instead employing a wind mass loss recipe to calculate it. We compute critical Carbon-Oxygen and Helium core masses to determine our lower limit to PI physics, and we provide two equations for $M_{\text{core}}$ and $M_{\text{final}}$ that can also be of use for binary population synthesis. Assuming the H envelope falls into the BH, we confirm the maximum BH mass below PI is $M_{\text{BH}} \simeq 93.3$ $M_{\odot}$. Our grid allows us to populate the traditional PI gap, and we conclude that the distribution of BHs above the traditional boundary is not solely due to the shape of the initial mass function (IMF), but also to the same stellar interior physics (i.e. mixing) that which sets the BH maximum.

\end{abstract}




\section{Introduction}

The outcomes from stellar evolution of the most massive stars are still highly uncertain, but could involve pair instability supernovae (PISNe) or black hole (BH) formation above and below the Pair Instability (PI) range. An accurate mapping of these outcomes is critical for our understanding of chemical evolution, as a full-fledged PI completely disrupts the entire star, enriching the environment, while a direct collapse to a BH implies no chemical enrichment at all \citep{Woosely02, Langer12, Yusof13, Limongi18, Higgins21, Volpato23}. The key factors deciding these outcomes involve the stellar core and envelope masses, with the former mostly set by the initial mass {($M_{\rm ZAMS}$)} and amount of overshooting, and the latter by metallicity-dependent stellar winds \citep{Vink21, Tanikawa21, Costa22}.
While direct observations of PISNe have been controversial \citep{Gal-Yam09, Schulze23}, a new window that could help address this question has opened up through the discovery of gravitational waves (GWs) from BH mergers by LIGO/VIRGO \citep{Abbott15}. While for a full analysis of BH masses from LIGO/VIRGO one would need to employ binary evolution and/or cluster dynamical simulations, specific features in the GW distribution, such as the maximum BH mass, are fundamental for all of massive star evolution, and not just for the small fraction of stars that eventually take part in a GW event.

Prior to the first GW event, most BH masses in our Milky Way had been determined from electromagnetic information yielding values up to just 10-20 $M_{\odot}$ {\citep[e.g.][]{Orosz11,Casares14}}. However the first GW event 150914 had components in the 30-40 $M_{\odot}$ range, which could most easily be explained by a reduced $Z$ content in a lower $Z$ galaxy {\citep[e.g.][]{Belczynski10,Spera15}} where wind mass loss from massive Wolf-Rayet (WR) stars driven by iron (Fe) ions would be expected to be lower \citep{Vink05}. Subsequent BH merger events also appear to "pile up" towards the {$\sim$35 $M_{\odot}$ mark \citep{Fishbach20,Kimball21,Wang22, Abbott23}}, which may be caused by the lower edge of PI, which for stars that have lost their hydrogen (H) envelope is thought to be located at $\sim$50 $M_{\odot}$ \citep{Woosley17, Farmer19, Renzo20_LowerPISN}. Note that pulsational PI (or PPI) may lower the maximum BH mass due to mass loss associated with pulses. So there are basically 2 lower edges to PI in the literature: one located in between direct BH formation versus the onset of PI physics, and a second one between PPI versus fully-fledged PISNe at higher masses. In this paper we focus on the former, with a boundary between direct BH collapse and the onset of PI physics, which turns out to be set by a critical core masses (for He and CO) that we consistently derive for objects with a large H envelope in this study.

A huge surprise came with the detection of BH masses above the traditional 50 $M_{\odot}$ stellar BH limit. In particular 
GW 190521 \citep{Abbott20_Detection} showed the existence of two heavy BHs of 66 $\left( ^{+17}_{-18} \right)$ and 85 $\left( ^{+21}_{-14}\right)$ $\ M_{\odot}$ merging into a product of 142 $\left(^{+28}_{-16}\right)$ $\ M_{\odot}$. One explanation for these high masses would be that these heavy BHs inside the traditional PI gap are themselves the product of earlier merging \citep{Fishbach20,fragione20,romero-shaw20, Renzo20_MergerScenarios}, but alternatively the assumptions in the stellar evolution considered up to that point in time was incorrect or incomplete. One possible solution would be that the nuclear reaction rates, in particular the $^{12}\mathrm{C}(\alpha,\gamma)^{16}\mathrm{O}$ nuclear reaction rate is uncertain \citep{2018ApJ...863..153T,Farmer20}, but even using standard nuclear reaction rates there might be a solution from a stellar evolutionary viewpoint. The key physical points are, on the one hand, to avoid entering the PI physics regime, maintain a sufficiently small core, and on the other hand to simultaneously retain a sufficiently large H envelope, both of which are possible for massive stars at reduced $Z$ \citep{Vink21}, where it was shown that a blue supergiant (BSG) can maintain a sufficiently large H envelope to produce an 85 $M_{\odot}$ heavy BH as long as the amount of core overshooting ($\alpha_{\rm ov} \leq 0.1$) is small (see the full motivation in Sect. \ref{ss-ParamSpace}).
Note that such a configuration of a small core and large Hydrogen envelope can also be produced in stellar mergers. \cite{Spera19} and \cite{DiCarlo19} found this to be a possibility using binary population synthesis, but single star models did not produce a BH over 60 $M_{\odot}$ in their set-up.

While we now know that such BSG models might potentially explain the mere existence of such heavy BHs {\citep[see also][]{Tanikawa21,Costa22}} in the traditional second mass gap (even without invoking mergers, second generation events, or nuclear physics adaptations) we still do not know the theoretical stellar upper BH mass limit before PI, nor do we know exactly which ZAMS mass would be able to produce this upper BH mass. Furthermore, while extreme 85 $M_{\odot}$ cases are exciting, due to the shape of the initial mass function (IMF) we would normally expect larger numbers of BHs to be found just above the 50 $M_{\odot}$ BH/PI boundary, than towards the 90 $M_{\odot}$ upper end. 

For these reasons, we set out to compute a significantly expanded (compared to \citet{Vink21}) grid of massive star models, to map the initial ZAMS masses into final masses (and related BH masses) due to stellar evolution assumptions of stellar interior mixing (core overshooting, rotation, semi-convection) and exterior stellar wind physics as a function of $Z$.
We first set out to determine the lower boundary of the PI region \citep{FowlerHoyle64,Barkat67,Woosley17,Farag22}. 
The upper boundary of the PI mass gap is estimated at $  M_{\text{ZAMS}} $$\sim 130$ $\ M_{\odot}$ \citep{Bond84,Heger03,Farag22}. Stars above this value are predicted to direct collapse to a black hole. However, as these stars are above the PI mass gap, they are not considered in this work.

The lower boundary of the PI mass gap, where PI physics takes place, had previously been estimated to occur at masses of $50 \ M_{\odot}$, such as \cite{Woosley17} or \cite{Farmer19} for pure helium stars.
The H-rich situation was explored in \cite{Vink21}, where a star of mass similar to the primary component of GW\,190521 retained almost all of its hydrogen (H) envelope into the final stages of its life, avoiding pair instability. 
The current work systematically continues the study of heavy BHs from stars in the traditional PI gap by exploring a much broader parameter space, allowing us not only to investigate which model parameters produce the heaviest BHs, but also how often this could occur, and what is the relative likelihood of the heaviest BHs in comparison to heavy BHs at the lower PI gap boundary. While these results are interesting for comparison to special features in GW data, isolated BHs could potentially also be tested by methods such as microlensing \citep{Lam22}.

Section \ref{s-Methods} explores our implementation of stellar parameters in MESA. Section \ref{s-CriticalCore} determines values for the critical core masses (herein referred to as the $M_{\rm crit}$ Experiment), while Section \ref{s-ParamSpace} explores the entire model grid and parameter space to verify which parameters produce the maximum BH mass (herein referred to as the Main Grid). In Section \ref{s-Mbh_Eqn}, we derive analytic fits to our data in the form of two equations. Section \ref{s-IMF} provides context to our work in terms of a BH distribution per initial mass. {Section \ref{s-discussion} provides discussion in the context of other work in the community, and} Section \ref{s-conclusions} presents our summarised conclusions. Appendix \ref{Ap-TableOfModels} presents a list of all 336 models and their input parameters, while Appendix \ref{Ap-ResTest} details the resolution test conducted before running the grid. Appendix \ref{Ap-RotationDiscussion} discusses the critical rotation rate in the context of informing future work, Appendix \ref{Ap-Winds} discusses our choice of mass loss rates, and Appendix \ref{Ap-MLT++_low_Z} discusses the effect of MLT++ at low metallicity and on the maximum black hole mass.

\section{Methods}
\label{s-Methods}

In this paper, we perform detailed stellar evolution employing a two-pronged strategy. First, we run a limited set of models all the way until \textcolor{black}{end of core oxygen burning} (unless they enter PI first) to establish whether the mass of the Carbon-Oxygen core (or He core) -- that sets the boundary between a model which undergoes PI and a model which does not (the \textit{critical} core mass) -- is constant across our parameter space (in terms of $\alpha_{\rm ov}$ etc.). This forms our $M_{\rm crit}$ experiment models.
If they are, then as a second step the critical core mass can be robustly employed to run a larger grid of 336 models just until the end of core Helium burning, as at this point both the Helium and Carbon-Oxygen cores will have been established. This forms our main grid.

The aim of this study is to derive the final masses, $M_{\rm final}$, from detailed stellar evolution models, and we do not study the 3D hydrodynamical collapse that creates a BH. For this latter part of BH formation we will rely on the work of others \citep[e.g.][]{Fernandez18}. 
In other words, our strategy is to derive accurate final masses ($M_{\rm final}$) from detailed stellar evolution modelling to provide a distribution on BH masses, and in particular on the maximum BH mass below PI. 
We will come back to this issue in the Discussion of Section \ref{s-discussion}.

\subsection{MESA Modelling}
\label{ss-MESAMod}

We use MESA version r15140 \citep{Paxton11, Paxton13, Paxton15, Paxton18, Paxton19} to evolve the main grid of 336 stellar evolution models from ZAMS to core Helium exhaustion, as well as a subset of models extending beyond Helium exhaustion and into core Oxygen burning (the $M_{\rm crit}$ Experiment), the details of which are further discussed in Section \ref{s-CriticalCore} and shown as part of the results in Section \ref{s-ParamSpace}. 

We include the standard Mixing Length Theory (MLT) of convection by \cite{Cox68} with an $\alpha_{\text{MLT}} = 1.82$ as in \cite{Choi16}. We employ the Ledoux criterion for convective stability with a varying efficiency parameter described in Section \ref{ss-ParamSpace}. During core He burning, massive stars evolve as supergiants with massive convective envelopes where radiative transport can become important. Supra-Eddington conditions can be realised locally inside the star in the regions of strong opacity bumps with structure models predicting strong inflation and density inverted layers on top to maintain hydro-static equilibrium. We note that the 1D treatment of radiative envelopes is an unsolved problem in Astrophysics \citep{Ishii99,Petrovic05,Graf12,Jiang15,Grassitelli21,Gilkis21,Klencki21,Romagnolo23} Some stellar models, such as BEC, allow inflation to occur \citep[e.g.][]{Koehler15}, while other models such as GENEC and MESA avoid the inflation and density inversion effects, either by swapping the density and pressure scale heights, enhancing mass loss in supra-Eddington layers \citep[e.g.][]{Ekstrom12}, or boosting convective energy transport efficiency employing MLT++ in MESA. 

The models of our main grid employ MLT++, as this is necessary for such high mass models to evolve into later stages of evolution (post-TAMS). This is required to avoid convergence issues in the envelopes resulting from density inversions in the outer layers where inefficient convective zones lead to radiation dominant regions. As for physical effects, MLT++ makes the star evolve hotter, and more compact. Calibration of the MLT++ parameters based on the red supergiant luminosity cutoffs in the Galaxy and the Magellanic Clouds is detailed in \cite{Sabhahit21}, while the description of how MLT++ is implemented is provided in \cite{Paxton13}.  For our grid, we use the default parameter values of MLT++ across 
the main grid models but showcase a model without MLT++ at very low metallicity in Appendix \ref{Ap-MLT++_low_Z}.

We include the effects of rotational mixing in our models, and therefore we implement the following instabilities; Solberg-Hoiland, Secular Shear Instability, Eddington-Sweet circulation and Goldreich-Schubert-Fricke, following \citep{Heger00}. The multiplicative diffusion factors to the diffusion coefficient of these different rotation-induced instabilities are all set to unity. The rotational component of the diffusion coefficient for mixing of material is multiplied by a factor $f_\mathrm{c} = 1/30$ (\textsc{am\_d\_mix\_factor} in MESA). The sensitivity of the rotation induced mixing to the composition gradient is controlled by the $f_\mu$ parameter which is set to 0.05 (\textsc{am\_gradmu\_factor} in MESA). Both these values follow \citet{Heger00}.

For our wind prescription, we follow \cite{Vink21} and modify the MESA "Dutch" wind, which is a collection of wind recipes from \cite{Vink01}, \cite{deJager88}, and \cite{NugisLamers00}. For our modified Dutch recipe we will use \cite{Vink01}, with a redefined Z dependence such that it is dependent on $Z_{\rm Fe}$ as opposed to the total metallicity in the $Z$ term. However, if the star drops below a certain temperature ($T_{\text{eff}} < 4 \text{kK}$), the model then switches to \cite{deJager88}. If the surface Hydrogen abundance drops below 0.4 with a surface temperature of $T_{\text{eff}} > 50 \text{kK}$, then the star is considered a Helium-rich Wolf-Rayet (WR) and uses the recipe of \cite{NugisLamers00}. 

As there are few observational constraints on mass-loss rates from massive stars in our parameter regime, there is notable uncertainty in stellar evolution modelling. For instance, a more appropriate WR recipe could be the recent theoretical one from \citep{SV20}, but as the bulk of our models remain cool, they do generally not enter this regime. More pressing questions would concern the absolute mass-loss rates during core H burning, where stars spend 90\% of their lives, and whether our models could encounter a phase of elevated mass-loss, due to becoming optically thick \citep{Vink2012}. Regarding the former, for canonical massive stars there is uncertainty in the mass-loss rates by a factor 2-3 \citep{Vink22,Bjorklund21,Krticka17}. However absolute mass-loss rates have only been calibrated at the higher $\sim80\,M_{\odot}$ range \citep{Vink2012}. Alternative recipes such as \citet{Bjorklund21} do not appear to provide the more appropriate higher rates in this transition regime, and we thus refrain from lowering mass-loss rates. Having noted this, as we are in the low $Z$ regime and H-burning mass-loss rates are modest, this choice would not affect the results.
Regarding the issue of mass-loss enhancement in close proximity to the Eddington limit \citep{Vink11}, we have recently included this in MESA models \citep{Sabhahit22,Sab23}, and as we show in Appendix\,\ref{Ap-Winds} and Fig.\,\ref{Fig-GauthamPlot}, the models in our grid are not in this regime. On the specific issue of luminous blue variable (LBV) mass loss \citep{Grassitelli21}, we refer the reader to Sect.\,4.1 in \citep{Vink21} for a discussion. In short, we cannot guarantee that LBV type mass loss would play some role, but it is unlikely due to the lower $Z$ \citep{Kalari18}.
As we estimated in \cite{Vink21} (see their Sect. 4.1), we could expect a total loss of $0.19\,M_\odot$ over the entire duration post core He-burning. This value is comparable to the amount of mass lost during core collapse and negligible in the context of our overall assumptions.
In any case, if mass loss was stronger (and/or envelope inflation more pronounced) then this could affect the Z-dependent cut-off on the maximum BH, as discussed in \cite{Vink21} and shown in Figure 3 of that paper. It would however have no effect on the BH maximum at low $Z$ itself, which is therefore a robust number in the mass-loss context, as discussed in \cite{Vink21}.

For the spatial meshing of the model during mass adjustment, the MESA code employs an inner Lagrangian meshing (constant $m$) and a homologous meshing (constant $q=m/M$) in the envelope \citep{Paxton11, Paxton15}. To facilitate a smoother run of our full model grid till end of evolution, we employ a deeper location for the switch from a Lagrangian meshing to a homologous meshing. 
A side effect of using such a deeper homologous mesh is that the models take larger timesteps during the main sequence ($dt \approx 10^4$ years as opposed to $ \approx 10^2$ without a deeper homologous mesh). Thus an additional timestep control limiting the model to a maximum timestep of $10^3$ years is used which allows us to properly resolve thermally unstable layers, mainly semiconvective layers. The usage of the maximum timestep control also has the effect of limiting timesteps during short phases beyond core He burning, which allows models in the critical core mass experiment (Section \ref{s-CriticalCore}) to complete oxygen burning. For further information on spatial and timestep meshing, see Appendix \ref{Ap-ResTest}.

\subsection{Parameter Space}
\label{ss-ParamSpace}

For our study we are interested in physical parameters which affect the evolution of the star at low $Z$. These are parameters which either have a significant effect on the core size, or affect the final mass of the star. For this we vary initial mass, initial metallicity up to Small Magellenic Cloud (SMC) $Z_{\text{ SMC }} = 1/5 Z_{\odot}$, core overshooting, rotation, and semiconvection. We do not vary the $^{12}$C($\alpha$,$\gamma$)$^{16}$O reaction rate, instead opting to maintain the standard rates for the course of our grid, as we wish to focus on stellar evolution uncertainties. A brief discussion on uncertain $^{12}$C($\alpha$,$\gamma$)$^{16}$O reaction rate can be found in Section \ref{s-discussion}.

The final mass from stellar modelling can be constructed by considering the stellar core versus the envelope: 

\begin{equation}
    \label{Eq-Mbh_Methods}
      M_{\text{final}}~= ~M_{\text{core}}(M_{\text{ZAMS}}, \alpha_{\text{ov}}, \Omega / \Omega_{\text{crit}}) ~+ ~M_{\text{envl}} 
\end{equation}

Where $M_{\rm core}$ is the mass of the star's core, and $M_{\rm env}$ is the mass of the star's envelope above the core. Both parameters are functions of our entire parameter space, however there are parameters which have a higher degree of impact to each part. $M_{\rm core}$ is predominantly a function of ZAMS mass, amount of core overshooting and rotation, while $M_{\rm final}$ is predominantly governed by $M_{\rm ZAMS}$ and $Z$ which affects $M_{\rm envl}$. In Section \ref{s-Mbh_Eqn}, we define explicit equations to these parameters and their relationship. For our grid of models, we focus on five parameters which can potentially drastically affect the evolution of a very massive star (we discuss other parameters in Section \ref{s-discussion}). 
These parameters are initial mass ($ { M_\mathrm{ZAMS} }$), metallicity ($Z$), overshooting ($\alpha_{\text{ov}}$), rotation ($\Omega / \Omega_{\text{crit}}$), and semiconvection ($\alpha_{\text{sc}}$).

Initial mass is the mass of the star at the Zero-Age Main Sequence (ZAMS) and is the first factor in determining the potential BH mass, as it provides an upper limit. Our mass range includes the values; 60, 80, 90, 100, 110, 120, 150 $M_{\odot}$.

 Initial metallicity refers to the abundance of elements other than Hydrogen and Helium, but crucially Iron, in the star at the ZAMS point, and is an important factor in the evolution of a massive star due to metallicity dependent winds. As we scale our abundances according to \cite{Grevesse98}, all values of metallicity we use are given as fractions of Solar metallicity ($Z_{\odot}$). We define abundances according to three initial mass fractions, $Z_{\odot} = 0.02$. The initial helium mass fraction in our models is evaluated as follows: $Y = Y_{\text{prim}} + (\Delta Y/\Delta Z)\times Z$ using MESA default values of $Y_{\text{prim}} = 0.24$ and $(\Delta Y/\Delta Z) = 2$. This leaves the hydrogen mass fraction ($X$) defined from unity, where $X + Y + Z = 1$. We also apply this metallicity value to the OPAL Opacity tables from \cite{Rogers02}. Our metallicity values range from \Z $ = 10^{-3}$ $ Z_{\odot}$ to $Z_{\text{SMC}} = 0.004$.

Convective boundary mixing (CBM) is a fluidic phenomenon which occurs above and below convective regions in stars \citep{Langer12, Anders23}. Regions of convection in stars are determined by the Schwarzschild equation, however fluid packets rising or falling to this boundary can still overshoot the boundary, thus creating a region of extra mixing above and below convective regions. As this is a 3-dimensional phenomonenon, it is not possible to fully represent this in 1-dimensional stellar evolution, however CBM can have a dramatic effect on the evolution of a model \citep{Higgins19, Vink21}. For our models, we use "step" overshooting. Step overshooting assumes a solid block of extra mixing, with the length or height predetermined by $\alpha_{\text{ov}}$ which is a fraction of the pressure scale height, $H_P$. 

Appropriate values to be used for CBM are highly uncertain. Up to 2010, most massive star evolution modellers employed relatively small values (of order \aOv $= 0.1$), but in order to accommodate the presence of B supergiants and an extended width of the main sequence, \cite{Vink10} suggested that overshooting values up to \aOv $= 0.5$ should be considered.
Such an increased amount of overshooting with stellar mass has also been employed by  \citet{2012MNRAS.427..127B,2016A&A...592A..15C,2017ApJ...849...18C,2018A&A...618A.177C,Scott21}.
However, this is only part of the story, as asteroseismology studies of B stars in the 10-20\,$M_{\odot}$ range \citet{Bowman20} indicate a roughly equal distribution for low ($\alpha_{\rm ov} \leq 0.1$, \citet{Dupret04}) and high ($\alpha_{\rm ov} \simeq 0.44$; \citet{Briquet07}) values. 
Future asteroseismology \citep{Aerts19} is needed to study CBM for higher mass objects.
The origin for what may be the root cause for the spread of CBM values per mass bin remains to be understood \citep{Costa19,Higgins23}. Meanwhile, our strategy in constructing stellar models has been to consider both low 
(\aOv $= 0.1$) and high (\aOv $= 0.5$) values for
massive OB stars \citep{Higgins19,Higgins23}.
For these reasons, our adopted values here are similar: \aOv $= 0.1, 0.3, 0.5$.
We limit overshooting to only occur above the core, the same value of which is implemented through to Carbon depletion for $M_{\rm crit}$ experiment models, or to Helium exhaustion for the main grid.

Stellar rotation is a powerful factor in massive star evolution (for a review of which, see \cite{Maeder00}), able to induce chemical mixing and lead to large amounts of mass loss if a star is supercritical such as in \cite{Meynet06}. Thus, rotation can also enhance the effects of, and also be affected by, other parameters such as $Z$-dependent winds. For our models, we use rotation as a fraction of the critical breakup speed ($\Omega / \Omega_{\text{crit}}$), as we want to maintain the same rotational effects between ZAMS masses. Our values of rotation in our main grid (Section \ref{s-ParamSpace}) are \OmOmC $ = 0.2, 0.4$ which corresponds to surface rotation rates of $\sim 150, 300 \text{ km s}^{-1}$ \citep[as motivated by observed rotation rates for O stars at low $Z$][]{Ramirez-Agudelo13,Sabin-Sanjulian17,Ramirez-Agudelo17,Ramachandran19}. 
For our $M_{\rm crit}$ experiment and our main grid, we do not use rotation enhanced mass loss \citep{Muller14}. However, in Appendix \ref{Ap-RotationDiscussion} we discuss the specific case of high rotation.

The last parameter we vary is semiconvective efficiency. Semiconvection, first discussed in \cite{Kato66}, applies to a region above the core of a star during the Main Sequence, where there is slow mixing of chemical elements.
We use the Ledoux criterion to determine regions of semiconvection and full convection. A region is convectively unstable if Equation \ref{Eq-Ledoux_Conv} is fulfilled, while it is semiconvective if Equation \ref{Eq-Ledoux_SemiConv} is fulfilled. 

\begin{equation}
    \label{Eq-Ledoux_Conv}
     \nabla_{\text{ad}} + \frac{\phi}{\delta} \nabla_{\mu} < \nabla
\end{equation}

\begin{equation}
    \label{Eq-Ledoux_SemiConv}
     \nabla_{\text{ad}} < \nabla < \nabla_{\text{ad}} + \frac{\phi}{\delta} \nabla_{\mu}
\end{equation}

where $\nabla$ is the temperature gradient, $\nabla_{\text{ad}}$ is the adiabatic temperature gradient, and $\nabla_{\mu}$ is the chemical gradient. $\phi$ and $\delta$ are the density gradients with respect to temperature and mean molecular weight respectively. 
Semiconvection can be seen above the core during H burning, as the core retreats, leaving behind said chemical gradient. However, the largest effect on a star's evolution from semiconvection happens during the post-MS phases, and can affect the temperature of the model, and the presence and duration of "blue loops" \citep{Langer12}.

\subsection{Pair Instability}
\label{ss-PI}

\begin{figure}
   \centerline{\hspace*{0.015\textwidth}
               \includegraphics[width=0.45\textwidth,clip=]{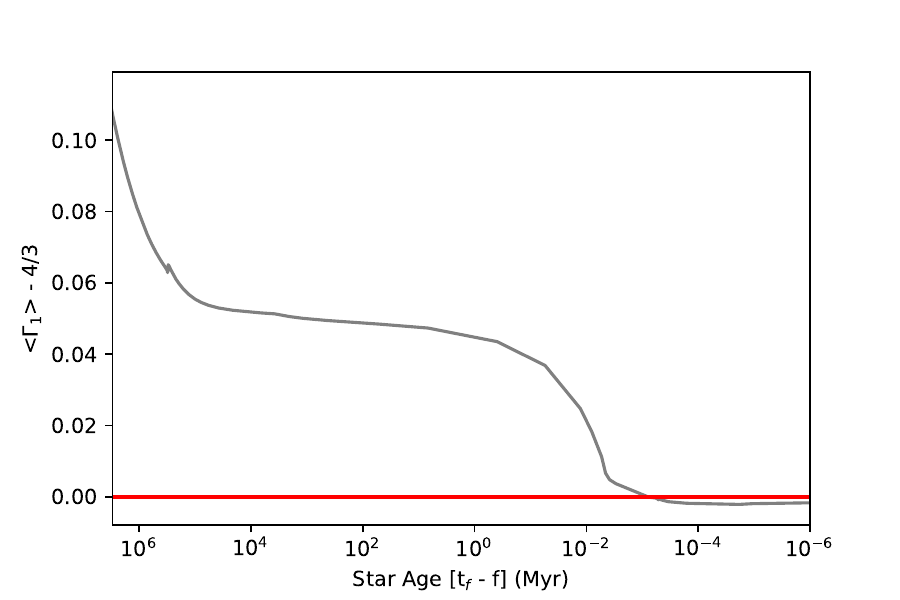}
              }
     \vspace{-0.35\textwidth}   
     \centerline{\Large       
         \hfill}
     \vspace{0.31\textwidth}
\caption{Evolution of a $95 M_{\odot}$ star in our grid with $\left< \Gamma_1 \right>$ explicitly calculated. The red line denotes where a star may become pair unstable by reducing below that point. The $\left< \Gamma_1 \right>$ is normalised to 0 in this case. {The horizontal axis follows the evolution of a star in logarithmic time until the end of the model. This allows us to see the evolution of $\left< \Gamma_1 \right>$ across the life of the star, and clearly show the last period of the star's life, where $\left< \Gamma_1 \right>$ drops below $4/3$.}}
\label{Fig-GammaIntegralTest}
\end{figure}

The PI region, from \cite{FowlerHoyle64}, is due to production of electron positron pairs in the core of a very massive star, leading to a reduction in the pressure component of the equation of state, and thus the gravitational contraction of the star. This leads to explosive oxygen burning, and can result in either a partial or complete explosion depending on the location of the instability. In the event of a partial explosion, large amounts of stellar material can be thrown off at once, forming a Pulsational Pair Instability Supernova (PPISN) \citep{Barkat67, Woosley17}. 
Alternatively, the star can be completely disrupted, and the resulting Pair Instability Supernova (PISN) leaves no compact remnant \citep{Barkat67, Farag22}.

The boundary for where PI or PPI occurs is a debated topic, for example \cite{Farmer19}, \cite{Costa22} and \cite{Volpato23}, as such several criteria are used to determine whether a star becomes pair unstable or not. Principally, the first criterion is that spontaneous pair production occurs in the star. For this to happen, the star needs to be massive enough such that the internal temperature reaches a point where a sufficiently large number of photons in the high energy tail of the Planck function have energies in excess of the rest mass. While this condition can be used as a diagnostic for PI, and can be visualised on a Temperature-Density diagram, alternatively one can instead use the size of Helium or Carbon-Oxygen cores as a determining factor. This has the benefit of providing a comparable diagnostic factor to determine the PI boundary.

For calculating which models became pair unstable, we calculated the average $\Gamma_1$ (see Equation \ref{Eq-int_gam1}) across the entire star at each timestep. As our current models do not include any implicit Hydrodynamics, we only split our model according to whether or not they became pair unstable at all ($\left< \Gamma_1 \right> < 4/3$).

An example of our calculation of $\left< \Gamma_1 \right>$ across the evolution of a stellar model is presented in Figure \ref{Fig-GammaIntegralTest}, showing a model which would undergo some form of Pair Instability.

The equation for $\Gamma_1$ is given in Equation \ref{Eq-gam1}. 

\begin{equation}
    \label{Eq-gam1}
    \Gamma_1 = \left( \frac{d \ln P}{d \ln \rho} \right)_{\rm ad}
\end{equation}

While it is possible to calculate the value of $\left< \Gamma_1 \right>$ for each model, it is also not necessary. As seen in \cite{Farmer19} and \cite{Farmer20}, it is possible to determine which models will undergo Pair Instability based on the {mass} of their Carbon-Oxygen or their Helium core mass ({see Section \ref{s-CriticalCore}}). As such, the majority of the grid can be categorised based on both of these criteria. For the $M_{\rm crit}$ experiment models which are used to calculate the critical core masses, some of these models do not deplete Oxygen, but all models are core Carbon depleted.
Similar to \cite{Farmer20, Costa21}, we use the following equation to calculate the average $\Gamma_1$ across our stellar model:
\begin{equation}
\label{Eq-int_gam1}
    \left< \Gamma_1 \right> = \frac{\int \Gamma_1 P d^3 r}{\int P d^3 r}
\end{equation}
 {from} \cite{Stothers99}, {who shows that this is a good approximation for determining the dynamical stability of the star. When the $\left< \Gamma_1 \right>$ of a star would fall below 4/3, PI would occur which causes the dynamical instability. As this is an approximation, this does not inform as to the location of the PI within the star, and as such only informs us that PI occurs, not whether the star would undergo Pulsational Pair Instability or full disruption (i.e. PISN).}

In the following Section \ref{s-CriticalCore}, we use this equation and stellar evolution models to demonstrate that the critical CO and Helium core masses are robust, and that therefore we use the core masses as our criteria for the PI boundary.

\section{$M_{\rm crit}$ Experiment}
\label{s-CriticalCore}

 The critical core mass is the mass of the stellar core such that the integral of the adiabatic index, $\left< \Gamma_1 \right>$, drops below 4/3. In this case, the model will undergo PI, such as in \citet{Farmer19}. This will be achieved with a critical mass for the Helium and Carbon-Oxygen cores, which has been shown in previous work to be an effective diagnostic for this purpose \citep{Farmer19,Costa21}. This section will provide values of the critical core masses with which we can apply to the main grid in Section \ref{s-ParamSpace} and to our population in Section \ref{s-IMF}.

 To do so, we use the model A1 ($M_{\rm ZAMS} = 90$ $M_{\odot}$, $Z = 1/10 \text{ th } Z_{\odot}$, \aOv$ = 0.1$, $\Omega / \Omega_{\rm crit} = 0.2$, $\alpha_{\rm sc} = 1$) from \citet{Vink21} as a starting point, as this is known to already be close to the BH/PI boundary. Then, we isolate one parameter ($M_\mathrm{ZAMS}$, $Z$, $\alpha_{\rm ov}$, $\Omega / \Omega_{\rm crit}$, $\alpha_{\text{sc}}$) which we vary, and evolve the models until Oxygen depletion. This provides an indication if the model undergoes PI, and thus allows us to map the PI limit for that parameter.
 
\begin{figure}
\begin{subfigure}{0.45\textwidth}
  \centering
  \includegraphics[width=\linewidth]{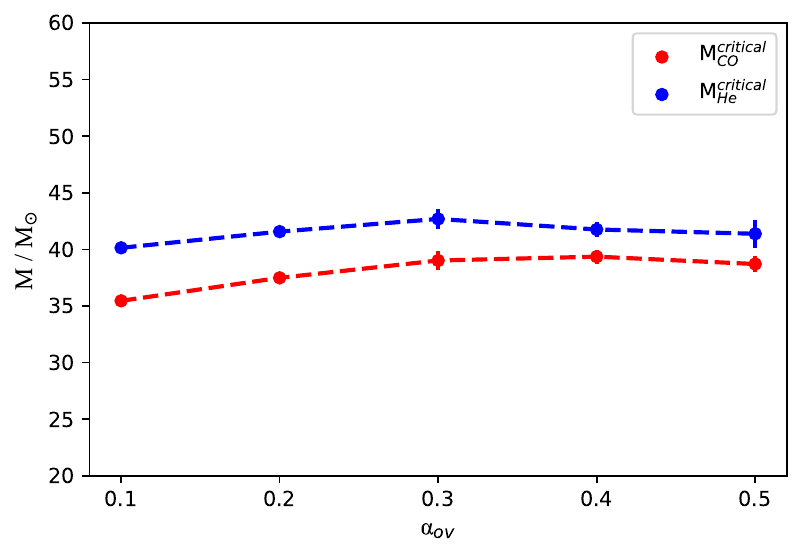}
\label{Fig-CriticalCoreMass_Ov}
\end{subfigure}
\begin{subfigure}{0.45\textwidth}
  \centering
  \includegraphics[width=\linewidth]{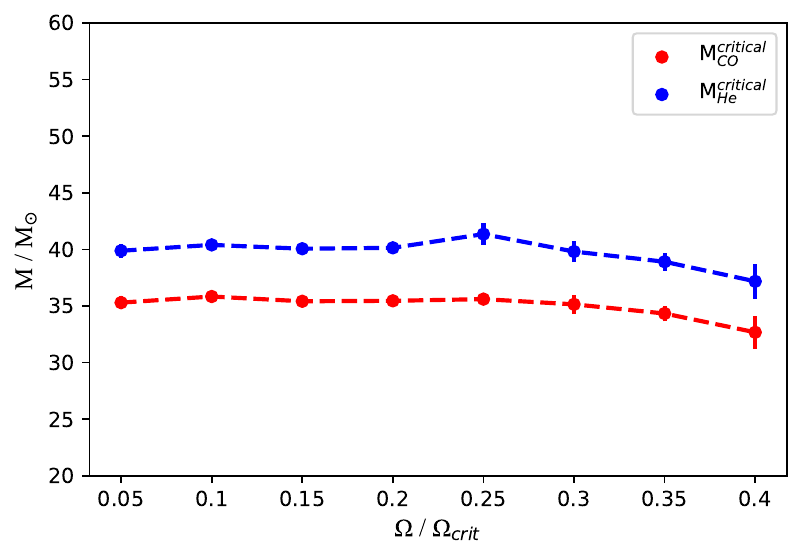}
\label{Fig-CriticalCoreMass_Rot}
\end{subfigure}
\caption{Critical core mass values for Helium ($M_{\text{He}}^{\text{critical}}$) and CO ($M_{\text{CO}}^{\text{critical}}$) when varying \aOv (top panel) and \OmOmC (bottom panel). Each data point represents the median between two models which straddle the PI boundary. The difference between these models' $M_{\text{CO}}$ and $M_{\text{He}}$ values provides us with an error bar, which is small enough to be hidden behind the data points on these graphs.}
\label{Fig-CriticalCoreMass_Ov_Rot}
\end{figure}

\begin{figure}
  \centering
  \includegraphics[width=1.05\linewidth]{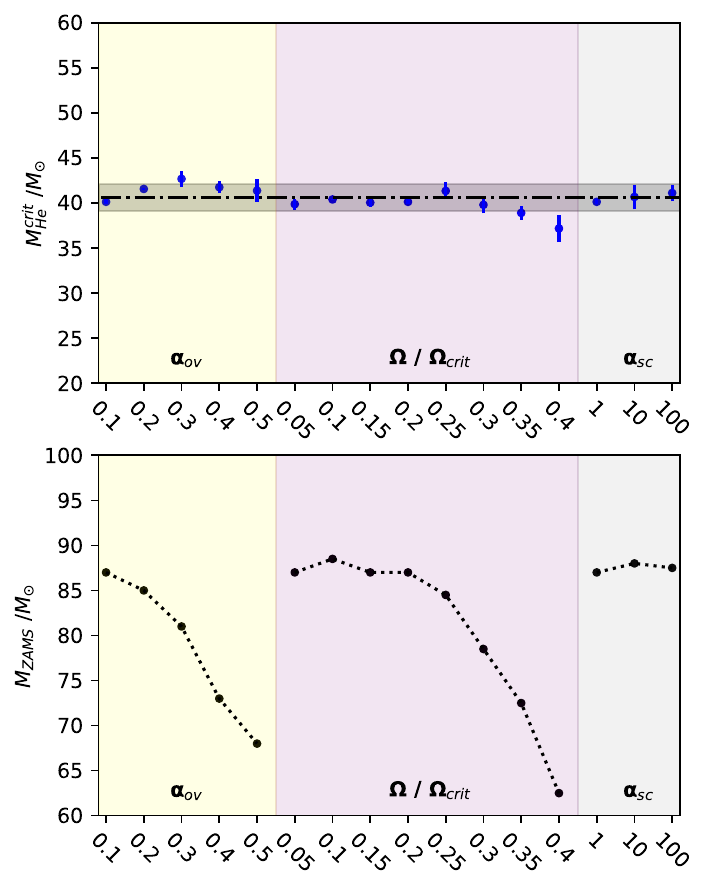}
\caption{The top panel shows the critical core mass for Helium varying across 3 parameters - $\alpha_{\rm ov}$, $\Omega / \Omega_{\rm crit}$, and $\alpha_{\rm sc}$. The average of these values - $40.6$ $M_{\odot}$ - is shown with the black dash-dot line, which we use as our fiducial critical Helium core mass ($M_{\text{He}}^{\text{critical}}$) throughout the paper. The black shaded region is our uncertainty, which represents the standard deviation of $M_{\text{He}}^{\text{critical}}$. The bottom panel shows corresponding ZAMS masses for the value of $M_{\text{He}}^{\text{critical}}$.}
\label{Fig-CriticalCoreMasses}
\end{figure}

In Figure \ref{Fig-CriticalCoreMass_Ov_Rot}, we show our calculated critical core mass across different values of overshooting (top panel) and rotation (bottom panel) for $ { M_{\text{CO}}^{\text{critical}} }$ and $ { M_{\text{He}}^{\text{critical}} }$. Stars that have core masses which are on or below their respective line will avoid Pair Instability, while those above undergo PI. 

 The ZAMS mass of the star corresponding to the critical core mass can be seen to vary in the yellow-shaded region of the bottom panel of Figure \ref{Fig-CriticalCoreMasses}. This figure combines the critical Helium core mass of Overshooting, Rotation and Semiconvection in the top panel, and provides the corresponding ZAMS masses in the bottom panel. As established, the critical core mass remains constant when varying $\alpha_{\rm ov}$, as this does not change the density of the core. Instead, increasing or decreasing \aOv has the effect of increasing or decreasing the total mass of the core. Thus, for the same corresponding Helium core mass, the ZAMS mass of the star will be inversely proportional to the amount of overshooting, $\alpha_{\rm ov}$.

 Regardless of variation in $M_{\text{ZAMS}}$ with $\alpha_{\text{ov}}$, $\Omega / \Omega_{\rm crit}$, or $\alpha_{\rm sc}$, the critical core mass is more or less constant. This {critical core mass} will be used in our determination of whether a model undergoes PI or not. 

By taking the average of each parameter's critical core mass boundary, and then averaging those values, we arrive at a critical CO core mass boundary of $ { M_{\text{CO}}^{\text{critical}} } =  36.3  \pm 1.8$ $M_{\odot}$ and critical Helium core mass boundary as $ { M_{\text{He}}^{\text{critical}} } =  40.6\pm 1.5$ $M_{\odot} $. Formal uncertainties are the combination of the standard deviations from the critical core masses in $\alpha_{\rm ov}$, $\Omega / \Omega_{\rm crit}$, and $\alpha_{\rm sc}$. The critical Helium core mass boundary is shown in the top panel of Figure \ref{Fig-CriticalCoreMasses}, where regions are shaded depending on varying parameter (yellow for overshooting, light pink for rotation and grey for semiconvection).

$\\$
To summarise, we have tested the role played by overshooting, semiconvection and rotation in setting the lower PI boundary limit by the critical core mass. We obtain the critical CO mass and critical He core mass of described in the above paragraph. More crucially, we find \textit{the critical core boundaries do not vary systematically with either overshooting, semiconvection, or rotation} within our parameter space. Such a critical core mass criterion remains independent of processes that can alter the core size and is a robust condition to determine the fates of our models.

\section{Main Model Grid}
\label{s-ParamSpace}

\begin{figure}
  \centering
  \includegraphics[width=1.1\linewidth]{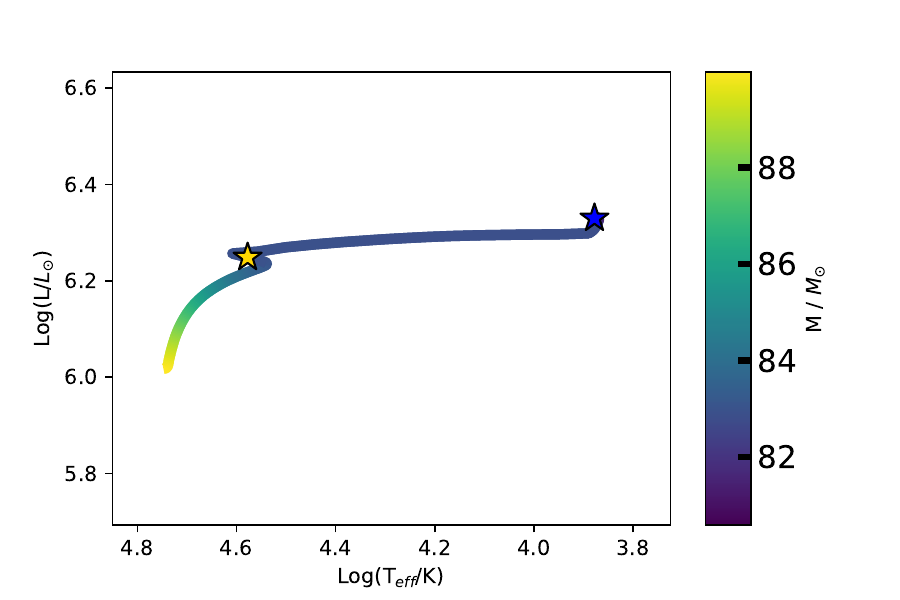}
\caption{ {Hertzprung-Russel diagram of a model with \Mi $= 90$ $M_{\odot}$, $Z = 1/10 \text{ th } Z_{\odot}$, \aOv $= 0.1$, \OmOmC $= 0.2$, and \aSC $= 1$. The gradient colour of the track illustrates the mass of the star. The gold and blue stars reflect the points of core Hydrogen and core Helium exhaustion respectively.}}
\label{Fig-HRD}
\end{figure}

\begin{figure}
   \centerline{\hspace*{0.015\textwidth}
               \includegraphics[width=0.45\textwidth,clip=]{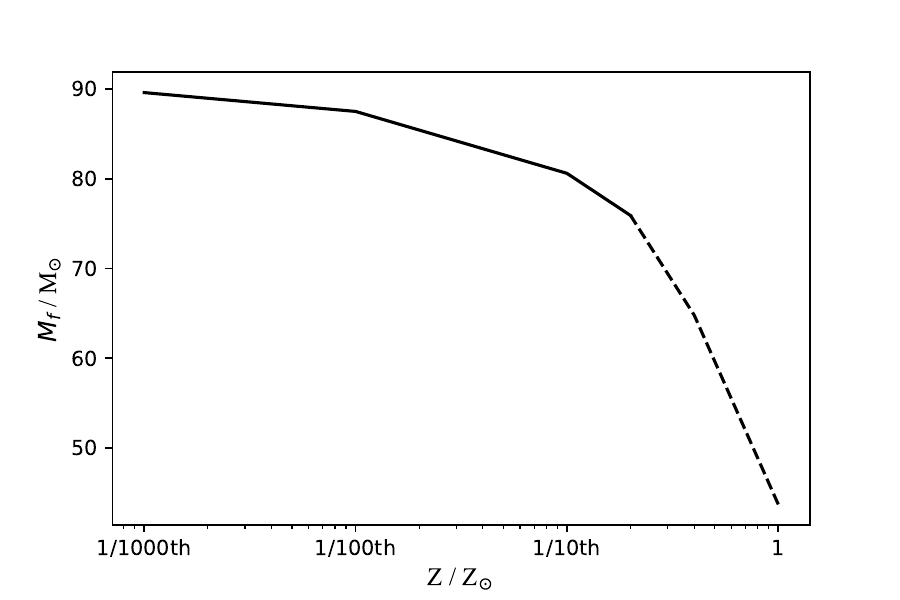}
              }
     \vspace{-0.35\textwidth}   
     \centerline{\Large       
         \hfill}
     \vspace{0.31\textwidth}
\caption{Final mass of models varying with metallicity. These models start with $90$ $M_{\odot}$ and vary from Solar metallicity to $1/1000 \text{ th }$Solar metallicity. Stars which are above the dotted line represents metallicities above that of the SMC, which are beyond our grid.}
\label{Fig-MetalsFinalMass}
\end{figure}

\begin{figure*}
\centering
\begin{tabular}{lc c c }

 & \cfbox{red}{$\alpha_{\text{sc}} = 1$} & \cfbox{red}{$\alpha_{\text{sc}} = 100$} \\
\rotatebox{90}{ \cfbox{blue}{$\alpha_{\text{ov}} = 0.1$} } & \includegraphics[width=0.45\linewidth,valign=m]{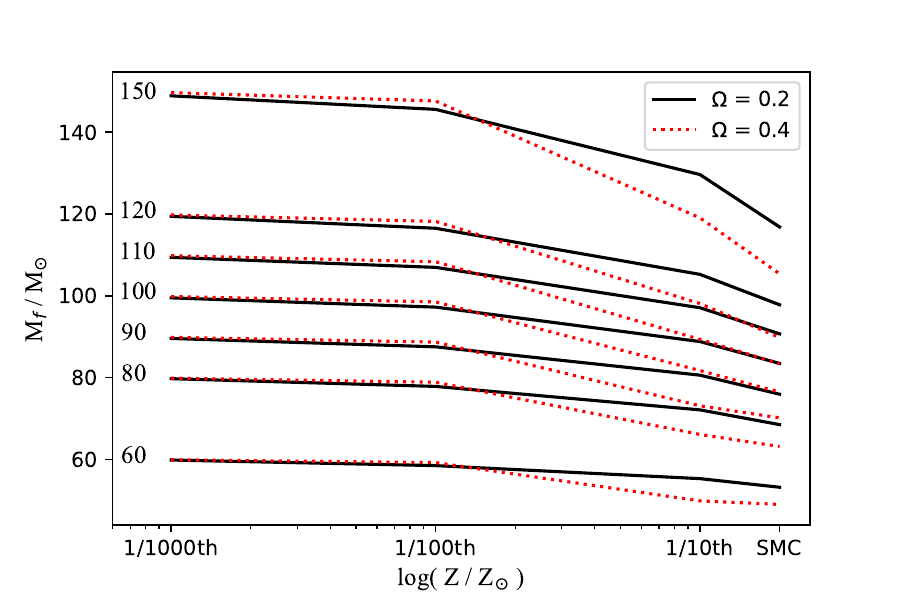} & \includegraphics[width=0.45\linewidth,valign=m]{mainGridMetV2_1sc_0.1Ov_allOm.pdf}\\
\rotatebox{90}{ \cfbox{blue}{$\alpha_{\text{ov}} = 0.3$} } & \includegraphics[width=0.45\linewidth,valign=m]{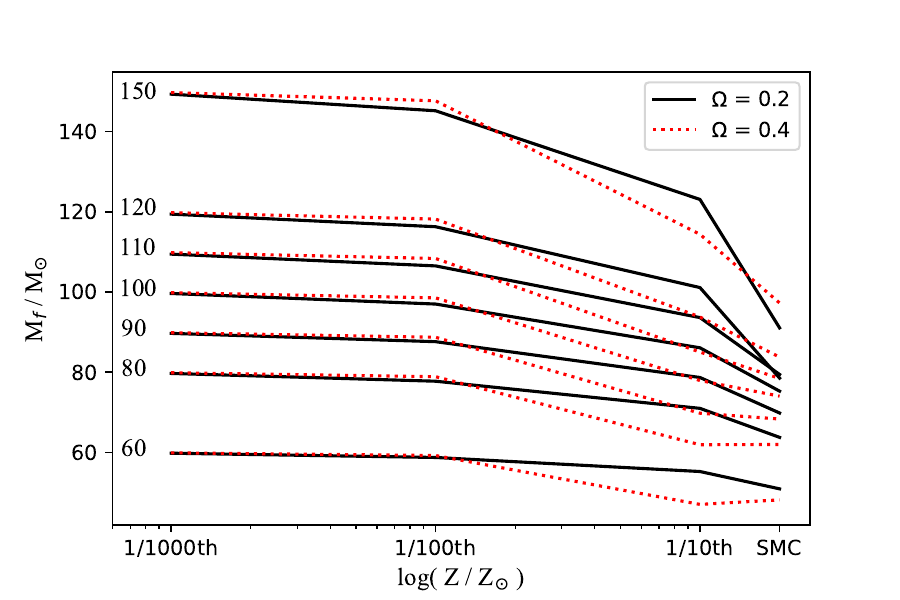} & \includegraphics[width=0.45\linewidth,valign=m]{mainGridMetV2_1sc_0.3Ov_allOm.pdf}\\
\rotatebox{90}{ \cfbox{blue}{$\alpha_{\text{ov}} = 0.5$} } & \includegraphics[width=0.45\linewidth,valign=m]{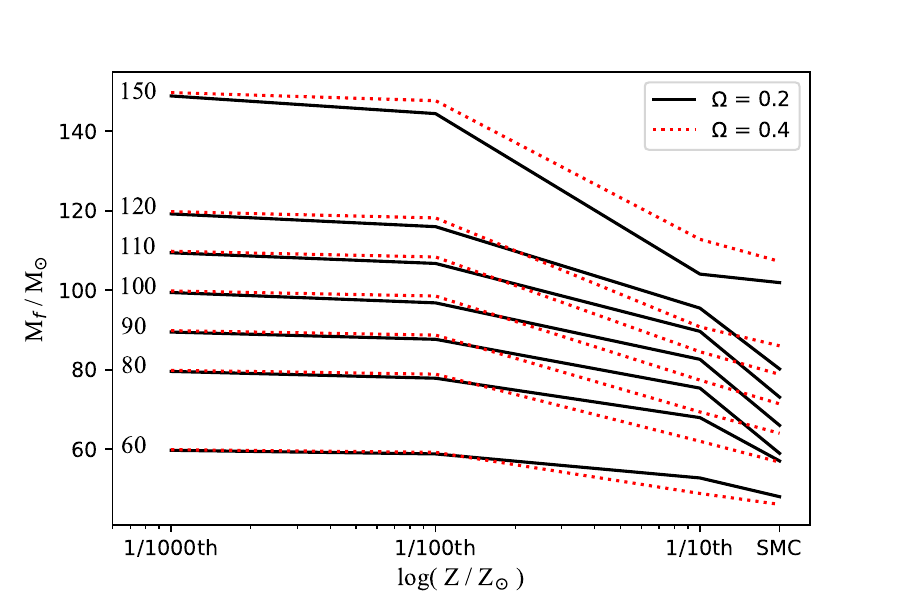} & \includegraphics[width=0.45\linewidth,valign=m]{mainGridMetV2_1sc_0.5Ov_allOm.pdf}\\
\end{tabular}
\caption{Final masses of stars as a function of metallicity. Low rotation (\OmOmC $= 0.2$) models are shown with the black solid line, while the higher rotation (\OmOmC $= 0.4$) models are represented with the red dotted line. Models with $\alpha_{\text{sc}} = 1$ are shown in the left column. Models with $\alpha_{\text{sc}} = 100$ are shown in the right column. Each row is divided by $\alpha_{\text{ov}}$ in descending order. Kinks in the curve can be found at SMC metallicity due to the presence of the bi-stability jump in the wind physics. The final mass of high rotation models at low metallicity are likely overestimated due to the lack of a supercritical rotation mass loss recipe.}
\label{Fig-MainGridMetModels}
\end{figure*}

In this section, we detail our grid of 336 models, describe the statistics of the grid, and discuss several features. Our parameter space is detailed in Table \ref{T-Params}. A breakdown of each model in the model grid can be found in Appendix \ref{Ap-TableOfModels}. The main grid of models was calculated to the end of Helium burning, and then the Carbon core mass taken and compared with the critical $ { M_{\text{CO}}^{\text{critical}} }$ from Section \ref{s-CriticalCore}. {An example of our models is shown in Figure \ref{Fig-HRD}, for a $M_{\rm ZAMS} = 90 M_{\odot}$ model.} 

\begin{table}
\centering
\begin{tabular}{c c c c c}
\hline
$ { M_\mathrm{ZAMS} }$ & $Z/Z_{\odot}$ & $\alpha_{\text{ov}}$ & $\Omega/\Omega_{\text{crit}}$ & $\alpha_{\text{sc}}$ \\
\hline
\hline
60 & 1/5 & 0.1 & 0.2 & 1 \\
80 & 1/10 & 0.3 & 0.4 & 100 \\
90 & 1/100 & 0.5 &  &  \\
100 & 1/1000 &  &  &  \\
110 &  &  &  &  \\
120 &  &  &  &  \\
150 &  &  &  &  \\
\hline
\end{tabular}
\caption{Detailed list of parameters used in our model grid. We provide the initial mass (\Mi), metallicity (\Z), overshooting parameter (\aOv), rotation rate as a function of critical rotation (\OmOmC), and semiconvective efficiency parameter (\aSC).}
\label{T-Params}
\end{table}

\subsection{Key Parameters of the Main Grid}
\label{ss-ParamSpace_KeyParams}

Metallicity is one of the key parameters of the main grid. Figure \ref{Fig-MetalsFinalMass} shows the effect of metallicity as a function of solar metallicity on the final mass of the star at the end of Helium burning. Below $1/10\text{th } Z_{\odot}$, the mass loss is negligible (less than 10 \% of total mass), however the function steepens as the models increase to SMC and Large Magellenic Cloud (LMC) metallicity, then finally $Z_{\odot}$ where the mass loss is non-negligible. 
A summary of the main grid of results is presented in Figure \ref{Fig-MainGridMetModels}. Immediately noticeable is the effect of metallicity on the final mass of a model. 
 For models at $Z_{SMC}$, {mass loss is non-negligible and is a significant ($\dot M_{\text{Total}} > 10 \% M_{\text{ZAMS}}$) percentage of the star's mass}. At $1/10\text{th } Z_{\odot}$, the mass loss becomes considerably weaker. At this metallicity, the stars retain the majority of their envelope through their evolution. This also means that the other parameters (such as overshooting, as shown in Section \ref{s-Mbh_Eqn}, and can be seen in the columns of Figure \ref{Fig-MainGridMetModels}) become a lot more important for determining the mass of a BH, as these parameters will affect the core size. For very low metallicities ($Z \leq 1/100 \text{th } Z_{\odot}$), the mass loss becomes negligible, with an average mass loss of $ < 2 \ M_{\odot}$.

As discussed in Section \ref{s-CriticalCore}, the core {mass} must be kept below \textcolor{black}{$   {M_{\text{CO}}^\mathrm{critical}} < 36.3 \pm 1.8 \ M_{\odot}$ or $   {M_{\text{He}}^\mathrm{critical}} < 40.6 \pm 1.5 \ M_{\odot}$} in order to avoid pair instability phenomena.

As can be seen in the bottom panel of Figure \ref{Fig-CriticalCoreMasses}, the ZAMS mass at which the a star would undergo PI is an inverse function of $\alpha_{\rm sc}$. See Section \ref{s-CriticalCore} for discussion on the interaction between $M_{\text{ZAMS}}$, $M_{\text{core}}^{\text{critical}}$, and \aOv. This indicates a strong correlation between these two parameters which is well defined in Section \ref{s-Mbh_Eqn}. For our values of rotation corresponding to ZAMS mass in the bottom panel Figure \ref{Fig-CriticalCoreMasses}, we see that larger values of $\Omega / \Omega_{\text{crit}}$ correspondingly decrease the maximum value of initial mass that a model can have before it becomes pair unstable.

\subsection{Features of Rotation and Semiconvection in the Main Grid}
\label{ss-ParamSpace_Rot_SC}

 {Rotation becomes important at very low metallicity ($Z \leq 1/100 \text{th } Z_{\odot}$). }At these metallicities, the wind mass loss is low enough such that models may spin up and reach critical breakup speed due to an inability to lose sufficient angular momentum during the main sequence. For models with which $Z = 1/100\text{th } Z_{\odot}$, and with initial $\Omega / \Omega_{\text{crit}} = 0.4$ reached critical breakup speed before core helium exhaustion, whereas models with initial $\Omega / \Omega_{\text{crit}} = 0.2$ would remain sub-critical during their evolution. However, at $1/1000\text{th } Z_{\odot}$, all models spun up to critical breakup speed in their lives. Conversely, at $1/10\text{th } Z_{\odot}$, all models were able to reach the end of core Helium burning without spinning up to breakup speeds. 

This is due to mass loss via winds being an important mechanism for dumping angular momentum from the star. For models with $Z \geq 1/10\text{th } Z_{\odot}$, the winds were able to eject enough angular momentum such that, even in the high rotation case, the stars did not reach breakup speeds. This relationship demonstrates a metallicity-dependent value of $\Omega / \Omega_{\text{crit}}$ which acts as a bifurcation point - values of rotation above this value critically spin up, and values below this are able to evolve without reaching critical rotation rates.

\begin{figure}
  \centering
  \includegraphics[width=\linewidth]{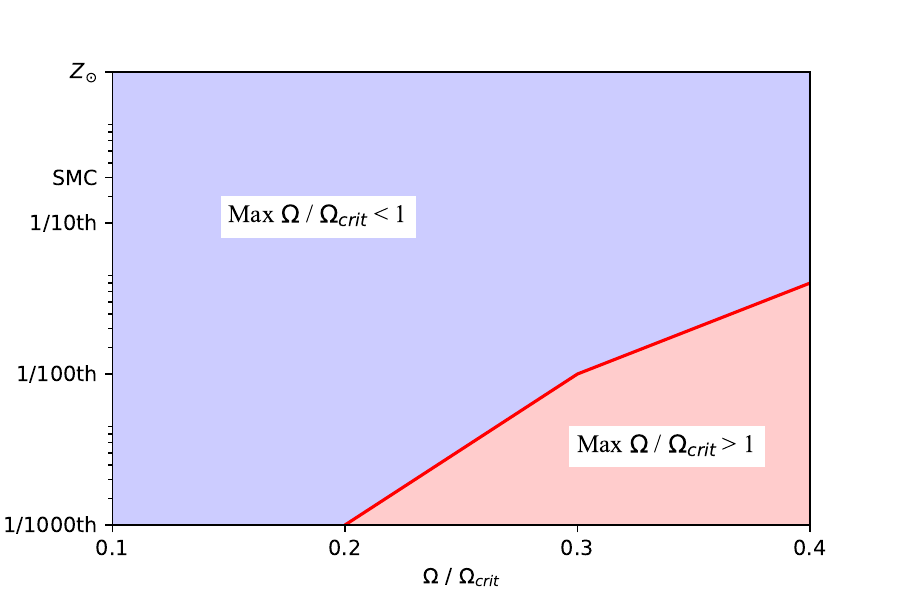}
\caption{Sketch of relationship between metallicity and maximum $\Omega / \Omega_{\text{crit}}$ {across the entire evolution}. Stars which have initial rotation values in the red region would not lose enough angular momentum via mass loss, leaving them to become supercritical when they contract and thus experience a period of very high mass loss. Above a certain metallicity, the initial rotation rate becomes irrelevant as the mass-loss rates are always substantial enough to shed enough angular momentum.}
\label{Fig-Rot_Z_sketch}
\end{figure}

A sketch of where this value lies is shown in Figure \ref{Fig-Rot_Z_sketch}, where the red region denotes a star which will spin up such that $\Omega/\Omega_{\text{crit}} > 1$, and therefore breakup. Whereas stars in the blue region are able to spin down or lose angular momentum such that their maximum rotation value remains below breakup speed. At SMC metallicity and greater, the strength of the mass loss would likely be enough for the star to lose most of its angular momentum regardless of initial rotation rate. It is interesting to note that, within our mass range, this feature is not mass dependent. 

This feature is important to the evolution of massive stars in the black hole/pair instability formation scenarios as this has a direct effect on the likelihood of formation for either scenario, as well as the mass distribution of black holes or PPISN/PISN progenitors. For models which have low rotation at the ZAMS, a relation exists where lower metallicity increases the maximum possible BH mass. However, by increasing rotation, this relationship changes due to large, episodic, mass loss at a cutoff metallicity - this cutoff is shown by the red line in Figure \ref{Fig-Rot_Z_sketch}. Below this cutoff, stars will experience episodes of rotationally induced mass loss which will lower their maximum BH mass, potentially on the order of tens of solar masses \citep{Meynet06}. Further discussion on high rotation at low metallicity is presented in Appendix \ref{Ap-RotationDiscussion}.

Overall, rotation can have various effects on the evolution and fate of a star. In addition to the episodic mass loss caused by rapid rotation as in \citet{Meynet06}, there is also the adjustment to the $\Gamma_1$ criterion as in \citet{Marchant20} (which has a noticeable effect from \OmOmC $\geq 0.3-0.4$), as well as chemically homogeneous evolution \citep{2005A&A...443..643Y}. The combined effects of these three rotational phenomena is beyond the scope of this Paper, and is a topic for future studies.

Semiconvection is the weakest parameter in this study.
At high values of initial mass, factors such as overshooting and rotation will dominate the effects of semiconvection to the point where models with weak or strong semiconvection are indistinguishable. At our lowest masses ($ { M_\mathrm{ZAMS} } = 60$ $M_{\odot}$), regions of semiconvection do become more prominent. However, this does not seem to affect the final fate of the model in any systematic way across our parameter space, except for isolated cases in CO core size.

For any indirect effect on mass loss, semiconvection was not noted to have a dominant effect here either. Temperature differences were typically small enough (nominally $< 0.1$ $dex$), or occurred for short timescales such that mass loss was almost the same between models of the same parameter space, and different $\alpha_{\text{sc}}$.

\begin{figure}
\begin{subfigure}{0.45\textwidth}
  \centering
  \includegraphics[width=\linewidth]{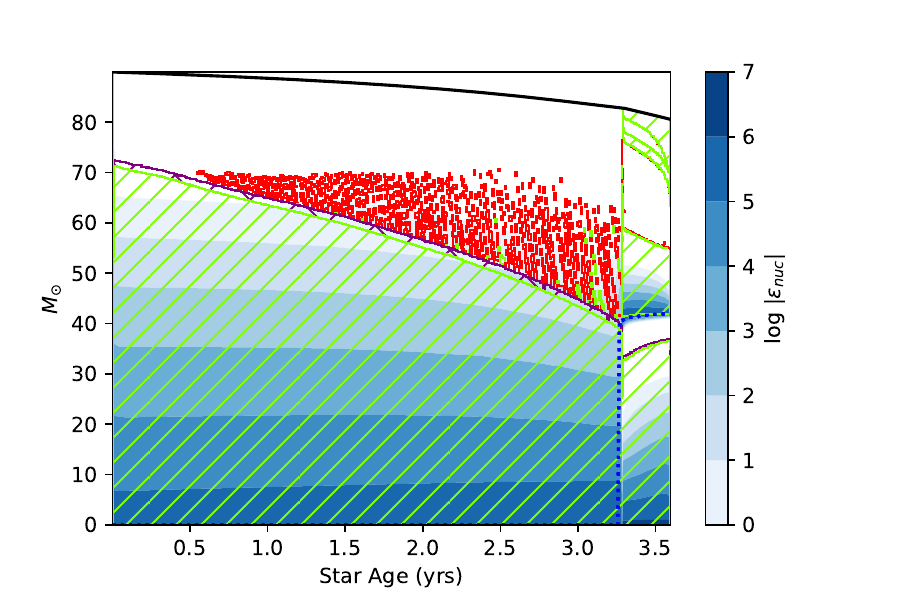}
  \caption{}
  \label{Fig-SC_Kipp_Low}
\end{subfigure}
\begin{subfigure}{0.45\textwidth}
  \centering
  \includegraphics[width=\linewidth]{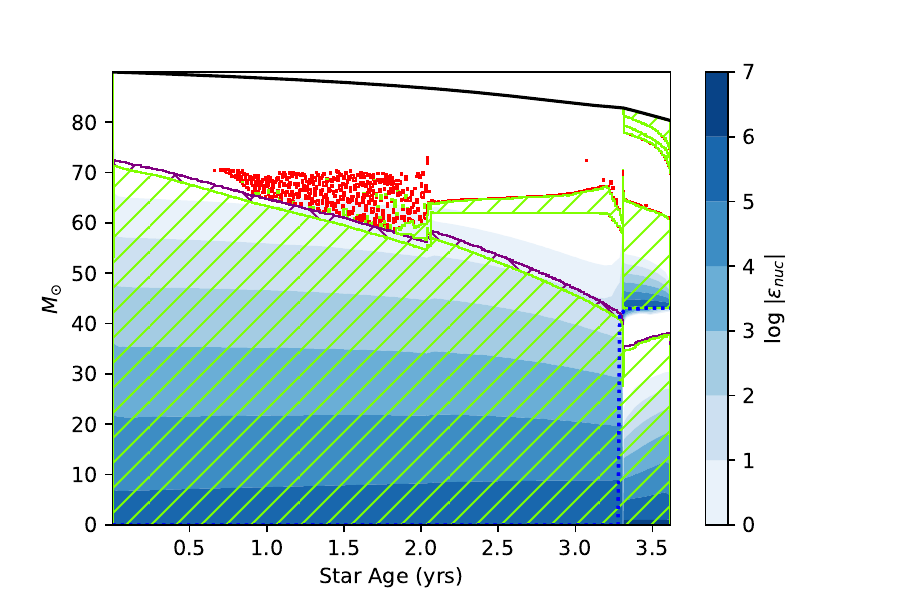}
  \caption{}
  \label{Fig-SC_Kipp_High}
\end{subfigure}
\caption{Stellar structure diagrams for a $90 \ M_{\odot}$ model with $\alpha_{\text{sc}} = 1$ (Figure \ref{Fig-SC_Kipp_Low}), and $\alpha_{\text{sc}} = 100$ (Figure \ref{Fig-SC_Kipp_High}). Fully convective regions are marked by green hash, while semiconvective regions are marked with red fill. Overshooting regions are seen in purple, and the Helium core mass is bounded by the blue dotted line. Blue shading indicates regions of nuclear burning.}
\label{Fig-SC_Kipps}
\end{figure}

Semiconvection does not directly affect the size of the core. The primary effect is instead related to the temperature of the model and this is mostly constrained to the post-TAMS environment, where longer blue-loops can occur for stars with $M_{\rm ZAMS} = 20-30 M_{\odot}$ \citep{Langer85}. In Figure \ref{Fig-SC_Kipps}, we can see the difference between high efficiency semiconvection for a $90$ $M_{\odot}$ model ($\alpha_{\text{sc}} = 100$, subfigure \ref{Fig-SC_Kipp_High}) and low efficiency ($\alpha_{\text{sc}} = 1$, subfigure \ref{Fig-SC_Kipp_Low}). The semiconvective efficiency shifted the TAMS temperature of the high efficiency model to be $0.1$ dex cooler. The red regions {in the Kippenhahn diagrams of Figure \ref{Fig-SC_Kipps}} denoting regions of semiconvective mixing become much more prominent with {lower} efficiency. This was not seen to have a significant effect on the evolution of the model, or the location of the PI limit.

\subsection{Statistics of the Main Grid}
\label{ss-RatesInitialMass}

All models of initial mass \textcolor{black}{$ { M_\mathrm{ZAMS} } \ge 90$ $M_{\odot}$} ended with core masses above our limit calculated in Section \ref{s-CriticalCore}. The exception is at $Z_{\rm SMC}$ where the line is just above the $90$ $M_{\odot}$ limit. As such, \textcolor{black}{almost }our entire model population of BH candidate models arrive from the initial mass range of \textcolor{black}{$M_{\rm ZAMS} = 60 - 80$ $M_\odot$}. While the rest of the models {undergo PI and therefore do not form part of our BH candidates}, they are valuable for examining trends, which factors into the {fits} of Section \ref{s-Mbh_Eqn}. Overall, \textcolor{black}{41} of our 336 models ended their evolution with CO cores below the predicted limit. The division of these parameters are described in the following paragraphs.

 \textcolor{black}{At $ { M_\mathrm{ZAMS} } = 80 \ M_{\odot}$ there are 13 models which had CO core masses below the PI limit.} The majority (9/13 models for this initial mass) of BH candidate models lost less than $10 \ M_{\odot}$ during their evolution, while a few more (4/13) ended up with final masses between $61-70 \ M_{\odot}$. These low final mass candidates were due to strong winds at $Z_{\rm SMC}$, with high rotation models losing approx $5 \ M_{\odot}$ more due to the {model's} resulting position in temperature space. Models with $ { M_{\rm ZAMS} } = 80 \ M_{\odot}$ at low metallicity ($1/100\text{th } Z_{\odot}$) were able to successfully remove enough angular momentum to avoid critical rotation.
Models at $ { M_\mathrm{ZAMS} } = 60 \ M_{\odot}$, had \textcolor{black}{35 models with CO core masses below the PI limit. Most of these models (21/35) had final masses above 50 $M_{\odot}$, while the rest had final masses between 46-60 $M_{\odot}$. The models with low masses were primarily contributed to by metallicities of 1/10 th $Z_{\odot}$ or $Z_{\text{SMC}}$.} Especially noticeable is that, at this initial mass, some (12/35) models of high overshooting ($\alpha_{\text{ov}} = 0.5$) were able to produce BH candidates in our applicable mass range (nominally, $M_{\text{final}} > 50 \ M_{\odot}$), which is not the case in the case where $M_{\text{ZAMS}} = 80 \ M_{\odot}$. 
$\\$
 {To conclude this section, we have shown that \aOv and \Mi are critical parameters for $M_{\text{core}}$, and that $Z$, \Mi, and \aOv are critical parameters for $M_{\text{f}}$, which we will use in the following section to parameterise our fits for the models. Additionally, we have found a metallicity dependent \OmOmC limit. This limit, which \textcolor{black}{can be seen sketched in Figure \ref{Fig-Rot_Z_sketch}, defines where a model would lose insufficient mass in order to remain below breakup speeds during the evolution. This was not found to be mass dependent in our range of ZAMS masses. }
This makes low metallicity models more susceptible to extreme mass loss, and thus lowering their potential black hole mass. Finally, we determine that semiconvection is unimportant within our parameter range in determining the final mass of a model, or its susceptibility to PI.}

\section{Numerical Fits of $M_\text{Core}$ and $M_{\text{final}}$}
\label{s-Mbh_Eqn}
 Based on our detailed stellar evolution models, we produce fits that may have use for population synthesis by ourselves or others.
As established in Section \ref{s-ParamSpace}, a set of clear relationships exist across the parameter space. Equation \ref{Eq-Mbh_Methods} gave $M_{\text{f}}$ as a function of core mass $M_{\text{core}}$ and envelope mass $M_{\text{envl}}$, which themselves are functions of our parameters. By fitting functions to these parameters, we can create a relation for the final mass.

Firstly, we look at the subset of our parameter space where $\Omega / \Omega_{\text{crit}} = 0.2$, and $\alpha_{\text{sc}} = 1$, as low rotating models are more likely to be stable and semiconvection is not a dominant variable (as per Section \ref{s-ParamSpace}). With this, we further define:
\begin{equation}
\label{Eq-Mcore_LowOm_LowAlphaSC_Definition}
    M_{\text{core}} = f (  { M_\mathrm{ZAMS} } , \alpha_{\text{ov}} ) , 
\end{equation}
\begin{equation}
\label{Eq-Mfinal_lowOm_LowAlphaSC_Definition}
    M_{\text{final}} = f (  { M_\mathrm{ZAMS} } , Z , \alpha_{ov} )
\end{equation}
By fitting the He core sizes to $ { M_\mathrm{ZAMS} }$ and $\alpha_{\text{ov}}$ based on 21 (7 $ { M_\mathrm{ZAMS} }$ by 3 $\alpha_{\text{ov}}$) models at constant $Z = 1/10 \text{ th } Z_{\odot}$, we achieve the following relation:

\begin{equation}
\begin{split}
     {M_\text{He core, fit} = -6.98 (\pm 1.6) + 0.51 (\pm 0.015) \cdot M_{\text{ZAMS}}} \\ {
     + 0.35 (\pm 0.19) \alpha_{\text{ov}} \cdot M_{\text{ZAMS}}^{1.03 (\pm 0.11)}}
\end{split}
\label{Eq-Mcore_LowOm_LowAlphaSC}
\end{equation}

which has a root mean square error of $0.82$ $M_{\odot}$. As mentioned, this relation was calculated for a constant metallicity value. However, when taking into account other metallicities, there was not a significant difference in the value of these coefficients, as metallicity does not have a significant effect on the core mass. The limit to this relation is approximately $Z < 1/100 \text{ th } Z_{\odot}$, for the same reasons described in Section \ref{ss-ParamSpace_Rot_SC}. 

 The mass of the convective core during the main sequence, which ultimately decides the mass of the formed Helium or Carbon-Oxygen core, is proportional to the ZAMS mass. In Equation \ref{Eq-Mcore_LowOm_LowAlphaSC}, we show that a linear relation well approximates the correlation between the core and ZAMS mass in the mass range considered in this study. This is because the core-to-total mass ratio increases with the ZAMS mass but then plateaus off as one approaches higher and higher ZAMS masses \citep[see][]{Yusof13, Sabhahit22}, which means the core and ZAMS mass relation approaches a linear trend. This might however break down for ZAMS masses beyond the mass range considered here. The main contributing factor towards such a departure from a linear relation is a switch to an optically thick wind, which is capable of eating away the core mass and significantly reducing it. However we note that such a switch to optically thick wind physics is metallicity dependent as shown in \citet{Sab23}, shifting to higher ZAMS masses towards lower $Z$. The model grid considered in this study largely avoids the optically-thick wind physics regime due to low $Z$ and the correlation between the core and ZAMS mass holds well.

Through fitting another set of 84 models, varying $\alpha_{\text{ov}}$, $ { M_\mathrm{ZAMS} }$ and $Z$ for the final mass of the star, $M_{\text{final}}$, we achieve: 

\begin{equation}
\begin{split}
    &  {M_{\text{final, fit}} = M_{\text{ZAMS}} \biggl[1 (\pm 0.006) \;\; - } \\ & { \bigg(0.02 (\pm 0.009)
    + 0.05 (\pm 0.019) \alpha_{\text{ov}} \bigg) \cdot M_{\text{ZAMS}}^{0.65 \pm 0.076} \cdot {\Bigg(\dfrac{Z}{Z_{\odot}}\Bigg)^{0.72 \pm 0.057}} \biggr]}
\end{split}
\label{Eq-Mfinal_lowOm_LowAlphaSC}
\end{equation}

\begin{figure}
\begin{subfigure}{0.45\textwidth}
  \centering
  \includegraphics[width=\linewidth]{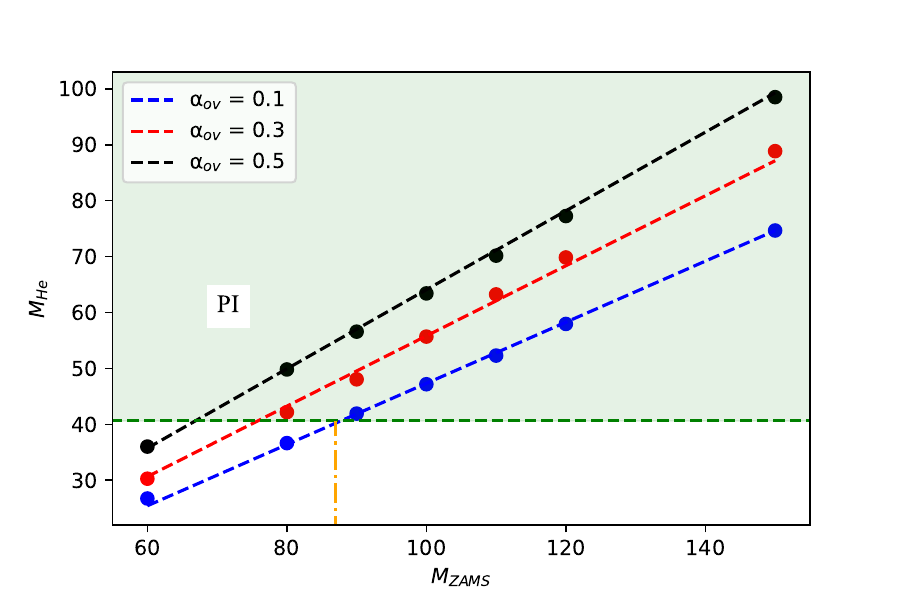}
  \caption{}
\label{Fig-Curve_Fit_Mcore} 
\end{subfigure}
\begin{subfigure}{0.45\textwidth}
  \centering
  \includegraphics[width=\linewidth]{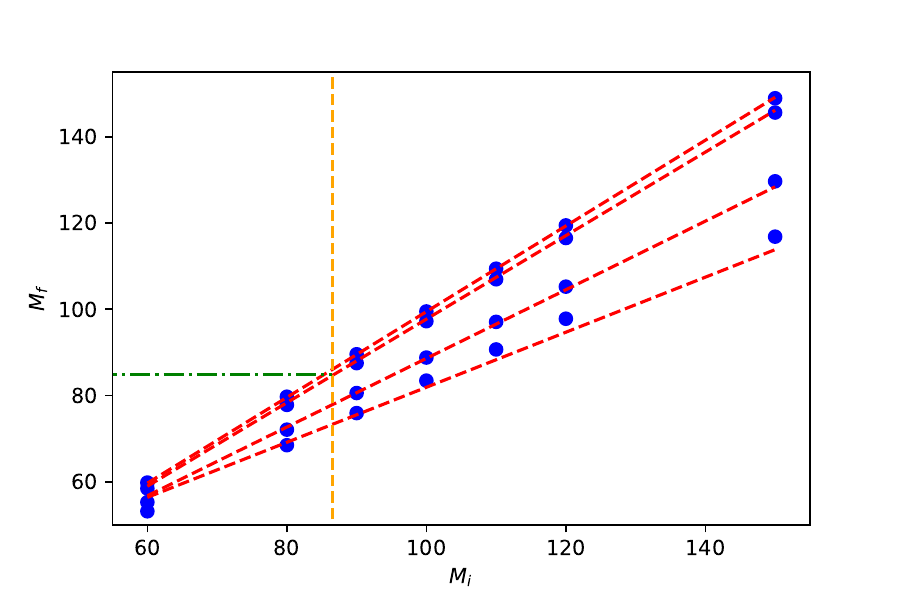}
  \caption{}
\label{Fig-Mi_Mf_fit}
\end{subfigure}
\caption{ \textcolor{black}{Subfigure \ref{Fig-Curve_Fit_Mcore} includes fitted lines for Helium core masses at constant $\Omega / \Omega_{\text{crit}} = 0.2$, metallicity $Z = 1/10 \text{ th } Z_{\odot}$, and semiconvection $\alpha_{\text{sc}} = 1$, from Equation \ref{Eq-Mcore_LowOm_LowAlphaSC}. The black, red and blue dots are data values from the models for overshooting values of $\alpha_{\text{ov}} = 0.5, 0.3, 0.1$ respectively, while the corresponding dashed lines are the fits for those values, and the green dashed line represents our calculated Helium core PI boundary $ { M_{\text{He}}^{\text{critical}} } = 40.6 \ M_{\odot}$. The greenshaded region represents where models would undergo PI. Subfigure \ref{Fig-Mi_Mf_fit} fits final masses as a function of initial mass and metallicity, with constant $\alpha_{\text{ov}} = 0.1$, $\Omega / \Omega_{\text{crit}} = 0.2$, and $\alpha_{\text{sc}} = 1$. Using the maximum initial mass in subfigure \ref{Fig-Curve_Fit_Mcore}, which is then marked in subfigure \ref{Fig-Mi_Mf_fit} by the orange line, the maximum final mass of the star is then correlated with the green line.}}
\end{figure}

This equation is accurate for $Z_{\rm SMC} < Z < 1/1000 \text{ th } Z_{\odot}$ and has a root mean square error of $3.09$ $M_{\odot}$. To find \textcolor{black}{any} BH mass \textcolor{black}{within our range }using our fits, first apply a \textcolor{black}{$40.6 \ M_{\odot}$} helium core mass limit, as discussed in Section \ref{s-CriticalCore} from which we can plot Equation \ref{Eq-Mcore_LowOm_LowAlphaSC} in Figure \ref{Fig-Curve_Fit_Mcore} and map the maximum initial mass before PI (x-axis) to the fits depending on a value of overshooting. It can be seen that, since the line fit trends to the left with increasing overshooting, that increasing overshooting has a negative correlation with the maximum initial mass before PI. We can do a similar mapping with Equation \ref{Eq-Mfinal_lowOm_LowAlphaSC}, plotted in Figure \ref{Fig-Mi_Mf_fit} this time for different values of metallicity. 

\textcolor{black}{In the limit of zero overshooting and metallicity, we obtain a maximum black hole mass of $M_{\rm BH} = 93.3$ $M_{\odot}$ which describes the physical maximum. Table \ref{T-IMF_Table_BH} shows no black holes within the $90-95$ $M_{\odot}$ range as the limit of our grid with \aOv is $0.1$ and \Z is $1/1000 \text{ th } Z_{\odot}$.}

\section{Populating the PI Gap}
\label{s-IMF}

\begin{table*}
    \centering
    \begin{tabular}{c | c | c c c | c }
    \hline
       $ { M_\mathrm{ZAMS} }$ & $N_{\text{bin}} / N_{\text{Total}}$ & $N_{\alpha{\text{ov}} = 0.1}$ & $N_{\alpha_{\text{ov}} = 0.3}$ & $N_{\alpha_{\text{ov} = 0.5}}$ & $N_{\text{BH}} / N_{\text{Total}}$\\
       \hline
       \hline 
       60-70  & 0.38 & 1/3     & 1/3   & 2/7  & 0.36   \\ 
       70-80  & 0.27 & 1/3     & 1/5   & 0    & 0.14   \\ 
       80-90  & 0.2  & 1/4     & 0     & 0    & 0.049  \\ 
       90-100 & 0.15 & 0       & 0     & 0    & 0      \\
        \hline
    \end{tabular}
    \caption{\textcolor{black}{Based on a population of 240,000 stars, this table shows the proportion of each initial mass bin ($ { M_\mathrm{ZAMS} }$) which forms black holes based on Equation \ref{Eq-Mcore_LowOm_LowAlphaSC}. The second column gives the number of stars in that mass bin as a fraction of the entire population - this is entirely based on the IMF. The next set of columns shows the ratio of models which form black holes under different values of overshooting (\aOv) for that population. The final column gives the number of black holes for that population as a fraction of the entire population of stars.}}
    \label{T-IMF_Table_Proportions}
\end{table*}

\begin{table*}
    \centering
    \begin{tabular}{ c | c | c c c | c c c }
    \hline
       $M_{\text{BH}}$ & $N_{\text{BH}}$ & $N_{\text{BH } \alpha{\text{ov}} = 0.1}$ & $N_{\text{BH } \alpha_{\text{ov}} = 0.3}$ & $N_{\text{BH } \alpha_{\text{ov} = 0.5}}$ & $N_{\text{BH SMC-10th}}$ & $N_{\text{BH 10th-100th}}$ & $N_{\text{BH 100th-1000th}}$ \\
       \hline
       \hline 
       60-65  & 40077 & 15948 & 13906 & 10221 & 5512  & 15845 & 18718\\
       65-70  & 26137 & 13420 & 10918 & 1799  & 3782  & 10383 & 11973\\
       70-75  & 16800 & 10314 & 6485  & 0     & 1970  & 6224  & 8606 \\
       75-80  & 8890  & 8363  & 526   & 0     & 797   & 3887  & 4207 \\
       80-85  & 5595  & 5595  & 0     & 0     & 0     & 2416  & 3179 \\
       85-90  & 967   & 967   & 0     & 0     & 0     & 0     & 967 \\
       90-95  & 0     & 0     & 0     & 0     & 0     & 0     & 0  \\
        \hline
    \end{tabular}
    \caption{\textcolor{black}{This table, using the same data and population of 240,000 stars as the previous (Table \ref{T-IMF_Table_Proportions}), bins the masses of the black holes in brackets of $5$ $M_{\odot}$. The second column gives the total absolute number of black holes for each mass bin. Columns 3-5 separate the effects of overshooting, with each column contributing black holes from one value of \aOv. Columns 6-8 split the models up based on metallicity, with each column showing the contribution of a range of metallicities to the total BH population.}}
    \label{T-IMF_Table_BH}
\end{table*}

\begin{figure*}
\begin{subfigure}{0.45\textwidth}
  \centering
  \includegraphics[width=1.02\linewidth]{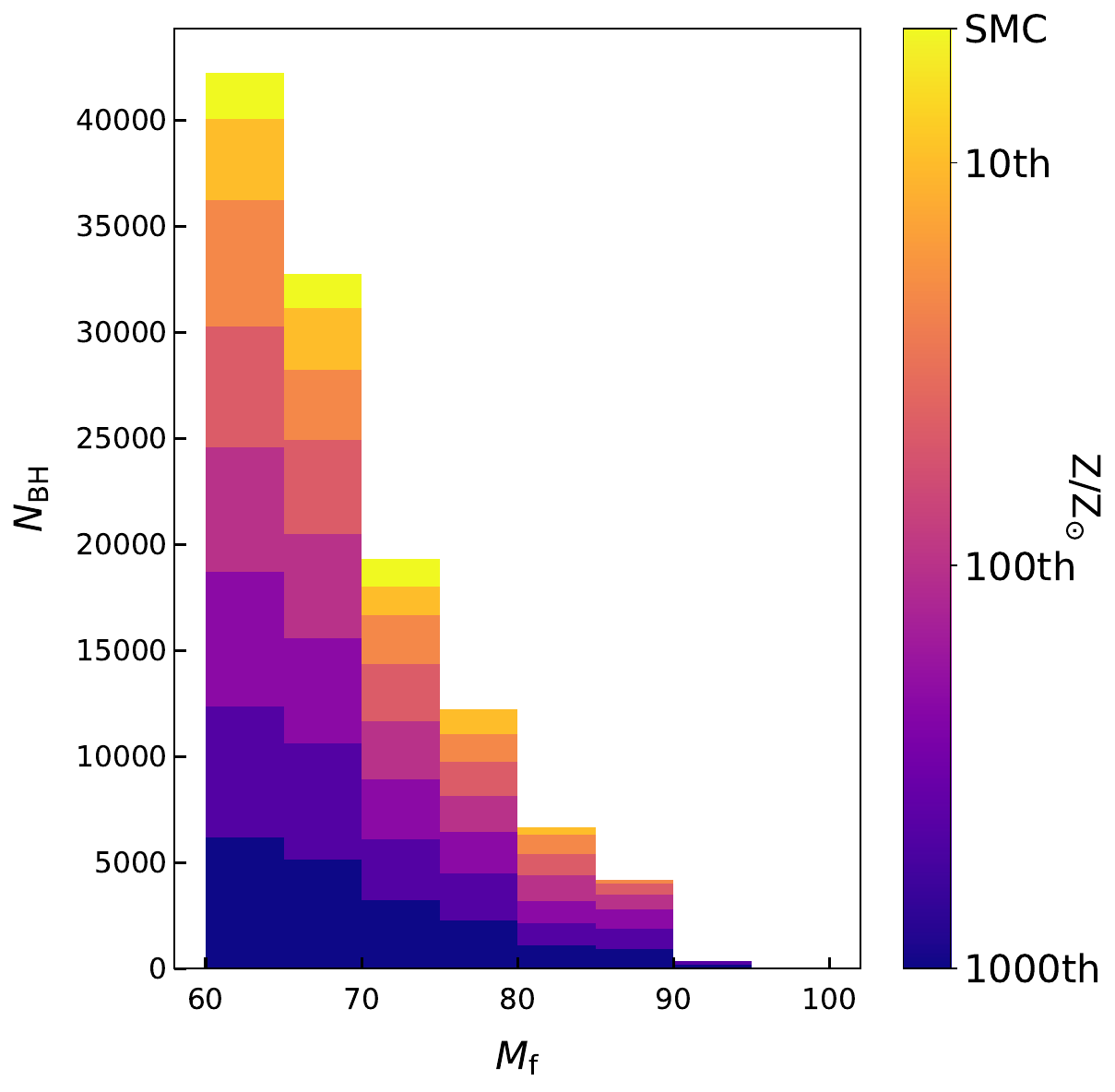}
  \caption{}
\label{Fig-BH_dist_Z}
\end{subfigure}
\begin{subfigure}{0.45\textwidth}
  \centering
  \includegraphics[width=0.98\linewidth]{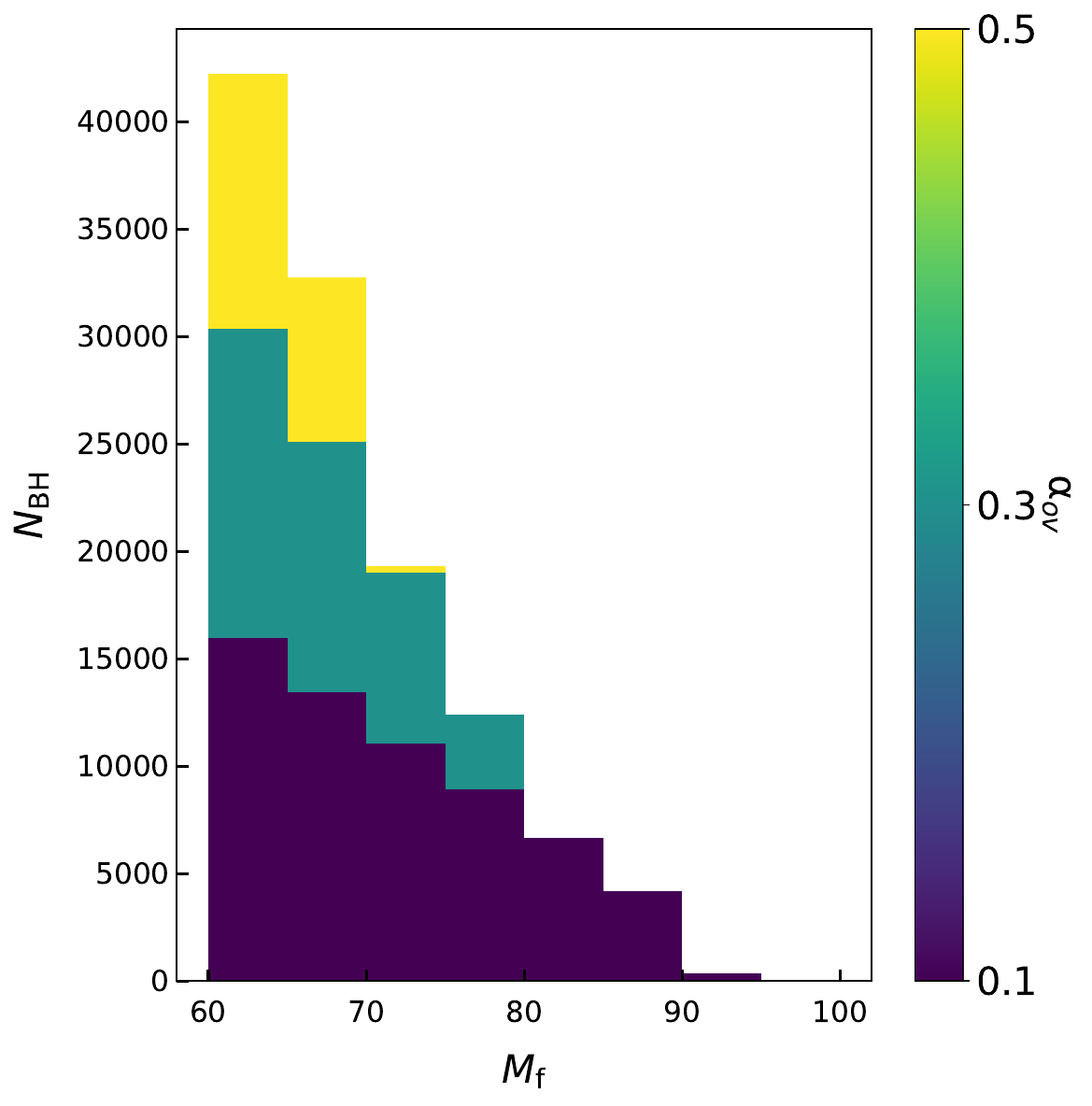}
  \caption{}
\label{Fig-BH_dist_ov}
\end{subfigure}
\caption{ \textcolor{black}{ Graphical representation of the numbers plotted in Table \ref{T-IMF_Table_BH}. Colours are arranged based on logarithmic metallicity in subfigure \ref{Fig-BH_dist_Z}, and \aOv in subfigure \ref{Fig-BH_dist_ov}. The x-axis represents each mass bin, while the y-axis is the total numbre of stars in that bin, coloured based on each parameter.}}
\end{figure*}

For this section, we populate the traditional PI mass gap ($> 50$ $M_{\odot}$) informed by the equations in Section \ref{s-Mbh_Eqn}. A sample of 240,000 stars - 10,000 stars for 3 values of overshooting and 8 metallicities - are populated between initial masses $60 < { M_\mathrm{ZAMS} } < 100$ $M_{\odot}$. Equal distribution on the log scale from $Z_{\rm SMC}$ to $1/1000 \text{ th } Z_{\odot}$ is assumed to obtain the metallicity values. Then we use Equation \ref{Eq-Mcore_LowOm_LowAlphaSC} to evaluate the He core size and check whether the model becomes pair unstable or not. Additionally, Equation \ref{Eq-Mfinal_lowOm_LowAlphaSC} is used to inform on the mass of the resulting BH. Tables \ref{T-IMF_Table_Proportions} and \ref{T-IMF_Table_BH} provide a detailed breakdown of our BH population, and weighs this against the Salpeter IMF ($M_{-2.35}$) for the population of 240,000 stars considered.

The 240,000 stars are initially divided into mass bins \textcolor{black}{$60-70, 70-80, 80-90, \text{ and } 90-100\; M_\odot$}  and the IMF-weighted fractions are shown in the second column of Table \ref{T-IMF_Table_Proportions}. The next three columns show the contributions from individual overshooting values. The final column shows the fraction of black holes formed corresponding to the initial mass range. This column takes into consideration of both the IMF weighting of the stars and the He core mass staying below \textcolor{black}{40.6 $M_\odot$} according to Equation \ref{Eq-Mcore_LowOm_LowAlphaSC}. In the low mass bin ($ { M_\mathrm{ZAMS} } = 60 - 70$ $M_{\odot}$), the three overshooting values contribute \textcolor{black}{almost} equally to the BH population, as the cores of these stars are too small regardless of the overshooting values in our range. As the initial mass range increases, we can see that the contribution towards heavy black holes becomes a strong function of overshooting. This is because as the initial mass increases, so does the core size. This strongly disfavours the higher overshooting values as a He core mass of \textcolor{black}{40.6 $M_\odot$} is considered our PI condition. In the \textcolor{black}{85-90} $M_\odot$ range, only a small fraction of the models can form heavy black holes, while the remaining go pair unstable. This shows that the production of heavy BHs in the PI gap, is not just a product of the IMF, but also of the amount of overshooting (extra mixing), showcased by the last column. 

In Table \ref{T-IMF_Table_BH}, we show the mass distribution of our IMF. All of the stars in this table have a He core mass less than \textcolor{black}{40.6 $M_\odot$}. The next three columns show the contribution from each overshooting value. A clear gradient is noticeable in the absolute BH numbers as a function of overshooting, which is a direct imprint from the previous table. Very heavy black holes with masses $M_{\text{BH}} > 80$ $M_{\odot}$ are only attainable at low overshooting. The maximum BH mass that can be attained with 0.3 overshooting is below $80$ $M_{\odot}$. This then forms a gradient of decreasing maximum BH mass as overshooting increases, culminating in a maximum BH mass of \textcolor{black}{$65-70$ $M_{\odot}$} at \aOv $= 0.5$. 

Additionally, the proportion of black holes also changes drastically depending on overshooting value, with higher overshooting resulting in fewer numbers of black holes in the same mass bin, as a correlative lower overshooting value. An exception to this is in the lowest mass bracket, where the progenitors have equal distributions across all values of overshooting. This is due to most of these black holes originating from the 60-70 $M_\odot$ initial mass range. 

The last three columns of Table \ref{T-IMF_Table_BH} give the absolute numbers distributed in three metallicity bins. We notice a gradient towards lower metallicities producing higher mass black holes, and in larger quantities. Similarly to high overshooting, the number of black holes drops off for higher mass bins. From our equations, the heaviest black hole is in the mass range $90-95$ $M_\odot$ and the required conditions are low overshooting \textcolor{black}{(\aOv $\leq 0.1$)} and low metallicity below $1/100 \text{ th } Z_{\odot}$.

Even though low overshooting and low metallicity are the most likely to produce the heaviest black holes, we can still see that large amounts of mixing can still result in BH production inside of the PI mass gap. This is counter-intuitive for such high overshooting, but encouraging, as it shows that the PI mass gap still \textit{shrinks} - even in the presence of larger amounts of CBM.

Finally, we quantify the ratio of low and high mass black holes within the PI mass gap, in an effort to help inform observations. For this, we will take the total number of black holes, across all metallicity and mixing values, and include only the values within the \textcolor{black}{$60-70$ $M_{\odot}$ and $80-90$ $M_{\odot}$} brackets, representing the lowest and highest mass black holes. We arrive at an approximate ratio of $10:1$, rounded to the nearest integer. 

\section{Discussion}
\label{s-discussion}

In this paper we have shown that the critical CO (or He) core mass does not depend on stellar modelling inputs such as core overshooting ($\alpha_{\rm ov}$), initial rotation rate ($\Omega$), or semi-convection ($\alpha_{\rm sc}$) (see upper panel Figure \ref{Fig-CriticalCoreMasses}). The final mass ($M_{\rm final}$) and also the black hole mass ($M_{\rm BH}$) do however depend on these inputs, but on some ($\alpha_{\rm ov}$) very strongly, and on others ($\alpha_{\rm sc}$) hardly at all in this regime. One could ask the question if there are any other parameters that we have not tested that could affect our results. Two notable aspects that have been discussed in recent literature are the $^{12}$C($\alpha$,$\gamma$)$^{16}$O reaction rate
\citep{Farmer19,Belc20single,Costa21} and envelope undershoot \citep{Costa21}. We discuss these two aspects in the next few paragraphs.

We first tested the effects of varying the $^{12}$C($\alpha$,$\gamma$)$^{16}$O reaction rate on the $M_{\rm{CO}}$ core mass near the PI gap. The thermonuclear reaction rates by \citet{Xu13} in NACRE II, which are implemented in the MESA code, are temperature-dependent and have upper and lower limits on the reaction for a given temperature range. As the uncertainty on the $^{12}$C($\alpha$,$\gamma$)$^{16}$O rate is not a simple gaussian, but changes non-linearly with temperature, we test the upper and lower limits of the nuclear reactions as provided by the experimental data in \citet{Xu13}.
At an initial ZAMS mass of 90 $M_{\odot}$ our stellar models \textcolor{black}{are at} the PI limit, and as such we test the upper and lower limits of the $^{12}$C($\alpha$,$\gamma$)$^{16}$O rate at the critical point. For this initial mass, the central temperature post-Helium burning ranges from 0.4-0.7 (10$^{9}$/K). We therefore adopt the lower and upper rates from \citet{Xu13} for this temperature range, which corresponds to a factor of 0.845 for the lower boundary and 1.16 for the upper boundary of the standard reaction rate. Our results show that the $M_{\rm{CO}}$ changes by at most 1-2 $M_{\odot}$ when adopting the lower or upper rates in this temperature range. 
Note that we have not systematically run critical $M_{\rm{CO}}$ experiments, as performed in for instance \cite{Farmer20} for H-poor models.

We next tested the effect of convective shell undershooting in our stellar models, with comparable $\alpha_{\rm{ov}}$ $=$ 0.1 during core H-burning and $\alpha_{\rm{under}}$ $=$ 0.1 on the H-shell during core He-burning, near the PI boundary with $M_{\rm{ZAMS}}$ $=$ 90 $M_{\odot}$. We find that the effect on the $M_{\rm{CO}}$ core is negligible with a difference of only 0.4 $M_{\odot}$ when compared to models with no undershooting. 
However, recent 3D hydrodynamic models \citep{Cristini16,Cristini19,Rizzuti23}, show that due to the stiffness of the lower boundary of convective shells, which means that the boundary is harder to penetrate, the convective undershoot value is found to be 1/5th of the upper overshoot value. By comparing the bulk Richardson's number in \citet{Rizzuti22} Table 1, we indeed see that the lower boundary is approximately 1/5th of the upper boundary for all 3D hydrodynamic models. Moreover, by comparing the velocities of the lower and upper convective boundaries in their Figure 3, we see that the velocities are significantly reduced at the lower boundary in comparison to the upper convective boundary, confirming that the convective mixing is reduced substantially below convective shells. See also \cite{Cristini16,Cristini19,Rizzuti23}.

So we have established that Equation \ref{Eq-Mfinal_lowOm_LowAlphaSC} may indeed represent an accurate representations of the final masses ($M_{\rm final}$) of our detailed stellar evolution modelling. However, what we have not yet discussed if our BH masses are accurately represented. As we mentioned at the start our strategy is to compute $M_{\rm final}$ and on the basis of hydrodynamical calculations of \citet{Fernandez18} we assume this to be linked to the final BH masses. There is one more important aspect of assumptions in our stellar evolution modelling to be discussed and that is how PPI might affect our results. As mentioned earlier, our lower PI boundary is the one between direct BH formation versus the onset of PI. It will only be the models towards the upper edge of the PI gap that will explode as PISNe, while those at lower masses may lose mass pulsationally and still form BHs. Figure 2 in \cite{Farmer19} is rather illustrative in this context. For models just above the BH/PI boundary it is shown that while PI pulses operate, this hardly affects the final masses. Only when models are comfortably above the BH/PI boundary (by roughly 10\,$M_{\odot}$) do pulses remove significant amounts of mass that could affect our BH mass distribution \citep[see also][]{Woosley21}{}. We will now estimate how PPI may affect our BH mass distribution. 

First, our BH distribution is dominated by ZAMS masses in the range 50-90\,$M_{\odot}$ while PI (including both PPI and PISNe) is considered to occur between 90 and 130  $M_{\odot}$. While we cannot exactly tell the mass boundary between PPI and PISNe, we assume that roughly half the objects in this 90-130\,$M_{\odot}$ range produce PISNe (thereby not contributing to BH formation) and the other half producing PPI. It is only this $\sim$1/2 of the 90-130\,$M_{\odot}$ mass bin that could affect our BH mass distribution. The second relevant realisation is that the IMF dictates that there are $\sim$3 times more stars in the mass bin between 50 and 90 $M_{\odot}$ than in the 90-130\,$M_{\odot}$ range. In other words, our BH mass distribution could be affected by about a factor $1/2 \times 1/3 = 1/6$, and is considered to be 83\% accurate. Future hydrodynamical PPI modelling is expected to improve our BH mass distribution.

In any case, despite this potential pollution, we can be confident that our BH mass 
provides meaningful insights into the true BH mass distribution for single stars. 
And while in the case of BH-BH merger events, binary evolution \citep{Belc16,Breivik16,Kruck18,Marchant19,Bavera20,Broek21} would need to be contemplated, especially if the bulk of BH-BH mergers would arise from the isolated binary scenario, we would expect binary evolution affects to mostly affect the lower range of our BH mass distribution. The upper part of our BH mass distribution, and especially the maximum BH mass, is not expected to depend on such additional complexities. 
In any case, even if we do not perform binary evolution, and cannot predict GW rates or the overall BH distribution originating from binary evolution, the specific observational BH maximum {\it feature} in the LIGO/Virgo data (of O4 and beyond) can be
tested in a meaningful way against our maximum BH mass from single star evolution.
Should the maximum (below PI) BH mass indeed turn out to be close to 93\,$M_{\odot}$ then this would provide confidence in our understanding of stellar evolution of the most massive stars. 
Should the maximum BH distribution appear notably different instead, then this would challenge our understanding of stellar evolution. One of the aspects that we would be particularly interested in testing further is that of $\alpha_{\rm ov}$ as this is the parameter that is empirically still highly uncertain, yet we found it to have the largest effect. 

Should even the uncertain stellar physics of CBM be resolved, but our predictions would differ from the GW features then this might indicate that the nuclear Astrophysics, and in particular the uncertain $^{12} C (\alpha, \gamma) ^{16}O$ rate \citep{2018ApJ...863..153T,Farmer20,Shen20,Farag22}, should be adapted.

\section{Conclusions}
\label{s-conclusions}

We have systematically investigated the Pair Instability region for very massive stars according to varying parameters of initial mass, overshooting, metallicity, rotation and semiconvection using the MESA stellar evolution code. In doing so, we have mapped out the effect of each of these parameters independently, and in relation to each other. This study also reduced the number of assumptions made about the formation of black holes in this region by considering the full evolution of the stars.

A summary of our conclusions is as follows:
\begin{itemize}
    \item We find the PI boundary at critical core masses of 
    $ {M_{\text{CO}}^{\text{critical}} } > 36.3 \pm 1.8\ M_{\odot}$ for Carbon-Oxygen $ { M_{\text{He}}^{\text{critical}} } > 40.6 \pm 1.5 \ M_{\odot}$ for Helium.
    \item The absolute maximum ZAMS mass for a first-generation BH in the PI boundary is \textcolor{black}{$\simeq 93.3 \ M_{\odot}$}, which \textcolor{black}{would also be the maximum BH mass in the physical limit}.
    \item While the upper limit of final mass is very strict in terms of appropriate parameters, stars of lower final mass are able to have higher values of rotation or overshooting. However rotation and overshooting remain the dominant parameters in limiting the maximum initial mass for BHs.
    \item We have analytically fit the core mass and final mass of the star to a pair of equations (Equations \ref{Eq-Mcore_LowOm_LowAlphaSC} and \ref{Eq-Mfinal_lowOm_LowAlphaSC}) which, together, provide a picture of the entire massive BH parameter space.
    \item As per our equations, overshooting is the dominant parameter for determining core size, aside from initial mass. Metallicity is the dominant parameter for determining envelope mass, and thus also prospective BH mass. 
    \item Rotation is also an important parameter, and is the only parameter which has a large effect with regards to both core mass and envelope mass. Additionally, rotation is able to influence the potential for a model to undergo supercritical mass loss, however this is also dependent on metallicity. Thus, high rotation (\OmOmC $> 0.4$) could stop BH production at low metallicity (\Z $< 1/100 \text{ th } Z_{\odot}$).
    \item Contrary to the case of lower mass blue supergiants \citep{Langer95,Schootemeijer19, Higgins20}, semiconvection is seen to be unimportant at our very high mass range, as it is completely overshadowed by overshooting and rotation. 
    \item \textcolor{black}{Lastly, we produce a population of BH candidates using our aforementioned equations to inform the distribution of stars. From this, we estimate that the ratio of producing a BH in the lower boundary of the mass gap ($60-70$ $M_{\odot}$) compared to the maximum ($80-90$ $M_{\odot}$) is approximately $10:1$}.
\end{itemize}

\section*{Acknowledgements}

The authors acknowledge MESA authors and developers for their continued revisions and public accessibility of the code. EW is funded by ST/W507925/1. JSV and ERH are supported by STFC funding under grant number ST/V000233/1. The authors would also like to thank the anonymous referee for their constructive comments which helped improve the presentation of the results.

\section*{Data Availability}

Input files for variables will be made publicly accessible via the MESA marketplace.



\bibliographystyle{mnras}
\bibliography{references} 



\appendix

\onecolumn
\section{Table of Models}
\label{Ap-TableOfModels}

Summary of all models run as part of the main grid. The first 5 columns describe the parameters of the models, while the 6th and 7th columns give the TAMS mass and Helium core mass. The last three columns provide the final mass, CO core and Helium core masses. Models which did not reach core helium exhaustion did not generate a CO core, and so their $M_{\rm CO}$ core property is left with a dash. Models that likewise did not reach TAMS due to supercritical rotation also have their TAMS columns properties ($M_{\rm TAMS}$, $M_{\rm He}$) replaced by a dash.

$\\$

\begin{tabular}{ p{1cm}p{1cm}p{1cm}p{1cm}p{1cm}|p{1cm} p{1cm}|p{1cm}p{1cm}p{1cm}}

\hline
\hline
 $Z$ & $M_\text{i}$ & \aOv & \OmOmC & $\alpha_{\text{sc}}$ & $M_{\text{TAMS}}$ & $M_{\text{He}}$ & $M_{\text{f}}$ & $M_{\text{CO}}$ & $M_{\text{He}}$ \\  
\hline

$10^{-3}$ & 60 & 0.1 & 0.2 & 1 & 59 & 27 & 59 & 23 & 27\\
$10^{-3}$ & 60 & 0.1 & 0.2 & 100 & 59 & 27 & 59 & 23 & 28\\
$10^{-3}$ & 60 & 0.1 & 0.4 & 1 & 59 & 53 & 59 & - & 53\\
$10^{-3}$ & 60 & 0.1 & 0.4 & 100 & 59 & 53 & 59 & - & 53\\
$10^{-3}$ & 60 & 0.3 & 0.2 & 1 & 59 & 29 & 59 & 27 & 30\\
$10^{-3}$ & 60 & 0.3 & 0.2 & 100 & 59 & 29 & 59 & - & 14\\
$10^{-3}$ & 60 & 0.3 & 0.4 & 1 & 59 & 55 & 59 & - & 55\\
$10^{-3}$ & 60 & 0.3 & 0.4 & 100 & 59 & 55 & 59 & - & 55\\
$10^{-3}$ & 60 & 0.5 & 0.2 & 1 & 59 & 33 & 59 & 31 & 34\\
$10^{-3}$ & 60 & 0.5 & 0.2 & 100 & 59 & 33 & 59 & 30 & 33\\
$10^{-3}$ & 60 & 0.5 & 0.4 & 1 & 59 & 56 & 59 & - & 56\\
$10^{-3}$ & 60 & 0.5 & 0.4 & 100 & 59 & 56 & 59 & - & 56\\
 \hline
$10^{-2}$ & 60 & 0.1 & 0.2 & 1 & 59 & 26 & 58 & 24 & 28\\
$10^{-2}$ & 60 & 0.1 & 0.2 & 100 & 59 & 30 & 58 & 29 & 33\\
$10^{-2}$ & 60 & 0.1 & 0.4 & 1 & 59 & 53 & 59 & - & 53\\
$10^{-2}$ & 60 & 0.1 & 0.4 & 100 & 59 & 53 & 59 & - & 53\\
$10^{-2}$ & 60 & 0.3 & 0.2 & 1 & 59 & 29 & 58 & 26 & 30\\
$10^{-2}$ & 60 & 0.3 & 0.2 & 100 & 59 & 30 & 58 & 25 & 28\\
$10^{-2}$ & 60 & 0.3 & 0.4 & 1 & 59 & 55 & 59 & - & 55\\
$10^{-2}$ & 60 & 0.3 & 0.4 & 100 & 59 & 55 & 59 & - & 55\\
$10^{-2}$ & 60 & 0.5 & 0.2 & 1 & 59 & 33 & 58 & 32 & 35\\
$10^{-2}$ & 60 & 0.5 & 0.2 & 100 & 59 & 33 & 58 & 32 & 35\\
$10^{-2}$ & 60 & 0.5 & 0.4 & 1 & 59 & 56 & 59 & - & 56\\
$10^{-2}$ & 60 & 0.5 & 0.4 & 100 & 59 & 56 & 59 & - & 56\\
 \hline 
$10^{-1}$ & 60 & 0.1 & 0.2 & 1 & 56 & 24 & 55 & 23 & 26\\
$10^{-1}$ & 60 & 0.1 & 0.2 & 100 & 56 & 26 & 55 & 23 & 27\\
$10^{-1}$ & 60 & 0.1 & 0.4 & 1 & 54 & 31 & 49 & 31 & 35\\
$10^{-1}$ & 60 & 0.1 & 0.4 & 100 & 54 & 31 & 47 & 31 & 35\\
$10^{-1}$ & 60 & 0.3 & 0.2 & 1 & 56 & 28 & 55 & 26 & 30\\
$10^{-1}$ & 60 & 0.3 & 0.2 & 100 & 56 & 28 & 55 & 25 & 29\\
$10^{-1}$ & 60 & 0.3 & 0.4 & 1 & 54 & 34 & 47 & 33 & 37\\
$10^{-1}$ & 60 & 0.3 & 0.4 & 100 & 54 & 34 & 48 & 33 & 37\\
$10^{-1}$ & 60 & 0.5 & 0.2 & 1 & 54 & 33 & 52 & 32 & 36\\
$10^{-1}$ & 60 & 0.5 & 0.2 & 100 & 54 & 33 & 52 & 32 & 36\\
$10^{-1}$ & 60 & 0.5 & 0.4 & 1 & 54 & 39 & 48 & 38 & 42\\
$10^{-1}$ & 60 & 0.5 & 0.4 & 100 & 54 & 39 & 48 & 38 & 42\\
 \hline
SMC & 60 & 0.1 & 0.2 & 1 & 54 & 23 & 53 & 21 & 25\\
SMC & 60 & 0.1 & 0.2 & 100 & 55 & 23 & 53 & 21 & 24\\
SMC & 60 & 0.1 & 0.4 & 1 & 52 & 27 & 48 & 26 & 30\\
SMC & 60 & 0.1 & 0.4 & 100 & 52 & 27 & 49 & 26 & 30\\
SMC & 60 & 0.3 & 0.2 & 1 & 53 & 28 & 50 & 28 & 31\\
SMC & 60 & 0.3 & 0.2 & 100 & 53 & 28 & 51 & 26 & 30\\
SMC & 60 & 0.3 & 0.4 & 1 & 51 & 31 & 48 & 30 & 34\\
SMC & 60 & 0.3 & 0.4 & 100 & 51 & 31 & 47 & 30 & 34\\
SMC & 60 & 0.5 & 0.2 & 1 & 51 & 33 & 48 & 32 & 35\\
SMC & 60 & 0.5 & 0.2 & 100 & 51 & 33 & 48 & 32 & 35\\
SMC & 60 & 0.5 & 0.4 & 1 & 50 & 35 & 46 & 35 & 38\\
SMC & 60 & 0.5 & 0.4 & 100 & 50 & 35 & 45 & 35 & 39\\
 \hline 
$10^{-3}$ & 80 & 0.1 & 0.2 & 1 & 79 & 38 & 79 & 33 & 38\\
$10^{-3}$ & 80 & 0.1 & 0.2 & 100 & 79 & 38 & 79 & 45 & 51\\

\end{tabular}
\newpage
\begin{tabular}{ p{1cm}p{1cm}p{1cm}p{1cm}p{1cm}|p{1cm} p{1cm}|p{1cm}p{1cm}p{1cm}}
\hline
\hline
 $Z$ & $M_\text{i}$ & \aOv & \OmOmC & $\alpha_{\text{sc}}$ & $M_{\text{TAMS}}$ & $M_{\text{He}}$ & $M_{\text{f}}$ & $M_{\text{CO}}$ & $M_{\text{He}}$ \\  
\hline

$10^{-3}$ & 80 & 0.1 & 0.4 & 1 & - & - & 79 & - & 72\\
$10^{-3}$ & 80 & 0.1 & 0.4 & 100 & - & - & 79 & - & 72\\
$10^{-3}$ & 80 & 0.3 & 0.2 & 1 & 79 & 41 & 79 & 36 & 39\\
$10^{-3}$ & 80 & 0.3 & 0.2 & 100 & 79 & 49 & 79 & 47 & 51\\
$10^{-3}$ & 80 & 0.3 & 0.4 & 1 & - & - & 79 & - & 74\\
$10^{-3}$ & 80 & 0.3 & 0.4 & 100 & - & - & 79 & - & 74\\
$10^{-3}$ & 80 & 0.5 & 0.2 & 1 & 79 & 47 & 79 & 45 & 49\\
$10^{-3}$ & 80 & 0.5 & 0.2 & 100 & 79 & 47 & 79 & 45 & 48\\
$10^{-3}$ & 80 & 0.5 & 0.4 & 1 & 79 & 75 & 79 & - & 75\\
$10^{-3}$ & 80 & 0.5 & 0.4 & 100 & 79 & 75 & 79 & - & 75\\
 \hline 
$10^{-2}$ & 80 & 0.1 & 0.2 & 1 & 78 & 37 & 77 & 34 & 39\\
$10^{-2}$ & 80 & 0.1 & 0.2 & 100 & 78 & 43 & 77 & 41 & 46\\
$10^{-2}$ & 80 & 0.1 & 0.4 & 1 & 78 & 72 & 78 & - & 72\\
$10^{-2}$ & 80 & 0.1 & 0.4 & 100 & 78 & 72 & 78 & - & 72\\
$10^{-2}$ & 80 & 0.3 & 0.2 & 1 & 78 & 42 & 77 & 37 & 41\\
$10^{-2}$ & 80 & 0.3 & 0.2 & 100 & 78 & 41 & 77 & 37 & 41\\
$10^{-2}$ & 80 & 0.3 & 0.4 & 1 & 78 & 74 & 78 & - & 74\\
$10^{-2}$ & 80 & 0.3 & 0.4 & 100 & 78 & 74 & 78 & - & 74\\
$10^{-2}$ & 80 & 0.5 & 0.2 & 1 & 78 & 47 & 77 & 45 & 49\\
$10^{-2}$ & 80 & 0.5 & 0.2 & 100 & 78 & 47 & 77 & 45 & 49\\
$10^{-2}$ & 80 & 0.5 & 0.4 & 1 & 78 & 75 & 78 & - & 75\\
$10^{-2}$ & 80 & 0.5 & 0.4 & 100 & 78 & 75 & 78 & - & 75\\
 \hline
$10^{-1}$ & 80 & 0.1 & 0.2 & 1 & 74 & 35 & 72 & 32 & 36\\
$10^{-1}$ & 80 & 0.1 & 0.2 & 100 & 74 & 36 & 71 & 33 & 37\\
$10^{-1}$ & 80 & 0.1 & 0.4 & 1 & 71 & 42 & 66 & 41 & 46\\
$10^{-1}$ & 80 & 0.1 & 0.4 & 100 & 71 & 42 & 65 & 41 & 46\\
$10^{-1}$ & 80 & 0.3 & 0.2 & 1 & 73 & 41 & 70 & 37 & 42\\
$10^{-1}$ & 80 & 0.3 & 0.2 & 100 & 73 & 41 & 70 & 36 & 40\\
$10^{-1}$ & 80 & 0.3 & 0.4 & 1 & 71 & 46 & 61 & 44 & 49\\
$10^{-1}$ & 80 & 0.3 & 0.4 & 100 & 71 & 46 & 62 & 43 & 48\\
$10^{-1}$ & 80 & 0.5 & 0.2 & 1 & 71 & 47 & 67 & 45 & 49\\
$10^{-1}$ & 80 & 0.5 & 0.2 & 100 & 71 & 47 & 67 & 45 & 49\\
$10^{-1}$ & 80 & 0.5 & 0.4 & 1 & 71 & 55 & 62 & 53 & 58\\
$10^{-1}$ & 80 & 0.5 & 0.4 & 100 & 71 & 55 & 64 & 53 & 58\\
 \hline
SMC & 80 & 0.1 & 0.2 & 1 & 71 & 33 & 68 & 31 & 35\\
SMC & 80 & 0.1 & 0.2 & 100 & 71 & 33 & 68 & 30 & 34\\
SMC & 80 & 0.1 & 0.4 & 1 & 67 & 38 & 63 & 37 & 41\\
SMC & 80 & 0.1 & 0.4 & 100 & 67 & 38 & 62 & 36 & 41\\
SMC & 80 & 0.3 & 0.2 & 1 & 67 & 40 & 63 & 40 & 43\\
SMC & 80 & 0.3 & 0.2 & 100 & 67 & 40 & 63 & 39 & 43\\
SMC & 80 & 0.3 & 0.4 & 1 & 67 & 43 & 61 & 42 & 46\\
SMC & 80 & 0.3 & 0.4 & 100 & 67 & 43 & 61 & 40 & 45\\
SMC & 80 & 0.5 & 0.2 & 1 & 63 & 46 & 57 & 45 & 48\\
SMC & 80 & 0.5 & 0.2 & 100 & 63 & 46 & 56 & 45 & 48\\
SMC & 80 & 0.5 & 0.4 & 1 & 64 & 50 & 56 & 49 & 53\\
SMC & 80 & 0.5 & 0.4 & 100 & 64 & 50 & 57 & 49 & 53\\
 \hline
$10^{-3}$ & 90 & 0.1 & 0.2 & 1 & 89 & 43 & 89 & 39 & 45\\
$10^{-3}$ & 90 & 0.1 & 0.2 & 100 & 89 & 46 & 89 & 42 & 48\\
$10^{-3}$ & 90 & 0.1 & 0.4 & 1 & 89 & 81 & 89 & - & 81\\
$10^{-3}$ & 90 & 0.1 & 0.4 & 100 & 89 & 81 & 89 & - & 81\\
$10^{-3}$ & 90 & 0.3 & 0.2 & 1 & 89 & 47 & 89 & 40 & 40\\
$10^{-3}$ & 90 & 0.3 & 0.2 & 100 & 89 & 48 & 89 & 41 & 41\\
$10^{-3}$ & 90 & 0.3 & 0.4 & 1 & - & - & 89 & - & 84\\
$10^{-3}$ & 90 & 0.3 & 0.4 & 100 & - & - & 89 & - & 84\\
$10^{-3}$ & 90 & 0.5 & 0.2 & 1 & 89 & 54 & 89 & 52 & 56\\
$10^{-3}$ & 90 & 0.5 & 0.2 & 100 & 89 & 54 & 89 & 46 & 46\\
$10^{-3}$ & 90 & 0.5 & 0.4 & 1 & 89 & 85 & 89 & - & 85\\

\end{tabular}
\newpage
\begin{tabular}{ p{1cm}p{1cm}p{1cm}p{1cm}p{1cm}|p{1cm} p{1cm}|p{1cm}p{1cm}p{1cm}}
\hline
\hline
 $Z$ & $M_\text{i}$ & \aOv & \OmOmC & $\alpha_{\text{sc}}$ & $M_{\text{TAMS}}$ & $M_{\text{He}}$ & $M_{\text{f}}$ & $M_{\text{CO}}$ & $M_{\text{He}}$ \\  
\hline

$10^{-3}$ & 90 & 0.5 & 0.4 & 100 & 89 & 85 & 89 & - & 85\\
\hline
$10^{-2}$ & 90 & 0.1 & 0.2 & 1 & 88 & 43 & 87 & 39 & 45\\
$10^{-2}$ & 90 & 0.1 & 0.2 & 100 & 88 & 46 & 87 & 41 & 47\\
$10^{-2}$ & 90 & 0.1 & 0.4 & 1 & 88 & 80 & 88 & - & 80\\
$10^{-2}$ & 90 & 0.1 & 0.4 & 100 & 88 & 80 & 88 & - & 80\\
$10^{-2}$ & 90 & 0.3 & 0.2 & 1 & 88 & 47 & 87 & 43 & 48\\
$10^{-2}$ & 90 & 0.3 & 0.2 & 100 & 88 & 48 & 87 & 41 & 44\\
$10^{-2}$ & 90 & 0.3 & 0.4 & 1 & 88 & 83 & 88 & - & 83\\
$10^{-2}$ & 90 & 0.3 & 0.4 & 100 & 88 & 83 & 88 & - & 83\\
$10^{-2}$ & 90 & 0.5 & 0.2 & 1 & 88 & 54 & 87 & 51 & 55\\
$10^{-2}$ & 90 & 0.5 & 0.2 & 100 & 88 & 54 & 87 & 52 & 56\\
$10^{-2}$ & 90 & 0.5 & 0.4 & 1 & 88 & 85 & 88 & - & 85\\
$10^{-2}$ & 90 & 0.5 & 0.4 & 100 & 88 & 85 & 88 & - & 85\\
 \hline 
$10^{-1}$ & 90 & 0.1 & 0.2 & 1 & 82 & 40 & 80 & 36 & 41\\
$10^{-1}$ & 90 & 0.1 & 0.2 & 100 & 82 & 42 & 80 & 38 & 43\\
$10^{-1}$ & 90 & 0.1 & 0.4 & 1 & 80 & 48 & 73 & 47 & 53\\
$10^{-1}$ & 90 & 0.1 & 0.4 & 100 & 80 & 50 & 70 & 48 & 53\\
$10^{-1}$ & 90 & 0.3 & 0.2 & 1 & 81 & 47 & 78 & 43 & 48\\
$10^{-1}$ & 90 & 0.3 & 0.2 & 100 & 81 & 47 & 78 & 43 & 47\\
$10^{-1}$ & 90 & 0.3 & 0.4 & 1 & 79 & 53 & 69 & 51 & 56\\
$10^{-1}$ & 90 & 0.3 & 0.4 & 100 & 79 & 53 & 70 & 50 & 56\\
$10^{-1}$ & 90 & 0.5 & 0.2 & 1 & 79 & 54 & 75 & 52 & 56\\
$10^{-1}$ & 90 & 0.5 & 0.2 & 100 & 79 & 54 & 75 & 52 & 56\\
$10^{-1}$ & 90 & 0.5 & 0.4 & 1 & 79 & 62 & 69 & 60 & 65\\
$10^{-1}$ & 90 & 0.5 & 0.4 & 100 & 79 & 62 & 69 & 60 & 65\\
 \hline 
SMC & 90 & 0.1 & 0.2 & 1 & 79 & 38 & 75 & 36 & 40\\
SMC & 90 & 0.1 & 0.2 & 100 & 79 & 39 & 75 & 35 & 39\\
SMC & 90 & 0.1 & 0.4 & 1 & 75 & 43 & 70 & 41 & 47\\
SMC & 90 & 0.1 & 0.4 & 100 & 75 & 43 & 69 & 41 & 46\\
SMC & 90 & 0.3 & 0.2 & 1 & 74 & 46 & 69 & 45 & 50\\
SMC & 90 & 0.3 & 0.2 & 100 & 74 & 46 & 69 & 45 & 49\\
SMC & 90 & 0.3 & 0.4 & 1 & 74 & 49 & 68 & 48 & 52\\
SMC & 90 & 0.3 & 0.4 & 100 & 74 & 49 & 66 & 46 & 51\\
SMC & 90 & 0.5 & 0.2 & 1 & 68 & 52 & 58 & 51 & 56\\
SMC & 90 & 0.5 & 0.2 & 100 & 67 & 52 & 59 & 51 & 56\\
SMC & 90 & 0.5 & 0.4 & 1 & 72 & 57 & 64 & 56 & 60\\
SMC & 90 & 0.5 & 0.4 & 100 & 72 & 57 & 64 & 56 & 60\\
 \hline 
$10^{-3}$ & 100 & 0.1 & 0.2 & 1 & 99 & 50 & 99 & 45 & 51\\
$10^{-3}$ & 100 & 0.1 & 0.2 & 100 & 99 & 51 & 99 & 44 & 50\\
$10^{-3}$ & 100 & 0.1 & 0.4 & 1 & - & - & 99 & - & 90\\
$10^{-3}$ & 100 & 0.1 & 0.4 & 100 & - & - & 99 & - & 90\\
$10^{-3}$ & 100 & 0.3 & 0.2 & 1 & 99 & 55 & 99 & 45 & 45\\
$10^{-3}$ & 100 & 0.3 & 0.2 & 100 & 99 & 54 & 99 & 43 & 43\\
$10^{-3}$ & 100 & 0.3 & 0.4 & 1 & - & - & 99 & - & 93\\
$10^{-3}$ & 100 & 0.3 & 0.4 & 100 & - & - & 99 & - & 93\\
$10^{-3}$ & 100 & 0.5 & 0.2 & 1 & 99 & 59 & 99 & 55 & 57\\
$10^{-3}$ & 100 & 0.5 & 0.2 & 100 & 99 & 61 & 99 & 58 & 63\\
$10^{-3}$ & 100 & 0.5 & 0.4 & 1 & - & - & 99 & - & 95\\
$10^{-3}$ & 100 & 0.5 & 0.4 & 100 & - & - & 99 & - & 95\\
 \hline 
$10^{-2}$ & 100 & 0.1 & 0.2 & 1 & 98 & 49 & 97 & 44 & 50\\
$10^{-2}$ & 100 & 0.1 & 0.2 & 100 & 98 & 53 & 97 & 47 & 54\\
$10^{-2}$ & 100 & 0.1 & 0.4 & 1 & 98 & 89 & 98 & - & 89\\
$10^{-2}$ & 100 & 0.1 & 0.4 & 100 & 98 & 89 & 98 & - & 89\\
$10^{-2}$ & 100 & 0.3 & 0.2 & 1 & 98 & 54 & 96 & 45 & 51\\
$10^{-2}$ & 100 & 0.3 & 0.2 & 100 & 98 & 55 & 97 & 45 & 45\\
$10^{-2}$ & 100 & 0.3 & 0.4 & 1 & 98 & 92 & 98 & - & 92\\

\end{tabular}
\newpage
\begin{tabular}{ p{1cm}p{1cm}p{1cm}p{1cm}p{1cm}|p{1cm} p{1cm}|p{1cm}p{1cm}p{1cm}}
\hline
\hline
 $Z$ & $M_\text{i}$ & \aOv & \OmOmC & $\alpha_{\text{sc}}$ & $M_{\text{TAMS}}$ & $M_{\text{He}}$ & $M_{\text{f}}$ & $M_{\text{CO}}$ & $M_{\text{He}}$ \\ 
\hline

$10^{-2}$ & 100 & 0.3 & 0.4 & 100 & 98 & 92 & 98 & - & 92\\
$10^{-2}$ & 100 & 0.5 & 0.2 & 1 & 97 & 61 & 96 & 59 & 63\\
$10^{-2}$ & 100 & 0.5 & 0.2 & 100 & 97 & 61 & 96 & 59 & 64\\
$10^{-2}$ & 100 & 0.5 & 0.4 & 1 & 98 & 94 & 98 & - & 94\\
 \hline 
$10^{-1}$ & 100 & 0.1 & 0.2 & 1 & 91 & 45 & 88 & 41 & 47\\
$10^{-1}$ & 100 & 0.1 & 0.2 & 100 & 91 & 49 & 88 & 44 & 50\\
$10^{-1}$ & 100 & 0.1 & 0.4 & 1 & 88 & 54 & 81 & 52 & 58\\
$10^{-1}$ & 100 & 0.1 & 0.4 & 100 & 88 & 54 & 80 & 51 & 57\\
$10^{-1}$ & 100 & 0.3 & 0.2 & 1 & 89 & 53 & 86 & 51 & 55\\
$10^{-1}$ & 100 & 0.3 & 0.2 & 100 & 89 & 53 & 86 & 51 & 55\\
$10^{-1}$ & 100 & 0.3 & 0.4 & 1 & 88 & 60 & 77 & 57 & 63\\
$10^{-1}$ & 100 & 0.3 & 0.4 & 100 & 88 & 60 & 78 & 57 & 63\\
$10^{-1}$ & 100 & 0.5 & 0.2 & 1 & 87 & 60 & 82 & 58 & 63\\
$10^{-1}$ & 100 & 0.5 & 0.2 & 100 & 87 & 60 & 82 & 58 & 63\\
$10^{-1}$ & 100 & 0.5 & 0.4 & 1 & 87 & 69 & 77 & 67 & 72\\
$10^{-1}$ & 100 & 0.5 & 0.4 & 100 & 87 & 69 & 77 & 67 & 72\\
 \hline 
SMC & 100 & 0.1 & 0.2 & 1 & 87 & 44 & 83 & 40 & 46\\
SMC & 100 & 0.1 & 0.2 & 100 & 87 & 44 & 83 & 39 & 45\\
SMC & 100 & 0.1 & 0.4 & 1 & 83 & 49 & 76 & 46 & 52\\
SMC & 100 & 0.1 & 0.4 & 100 & 83 & 49 & 76 & 46 & 52\\
SMC & 100 & 0.3 & 0.2 & 1 & 81 & 52 & 75 & 51 & 56\\
SMC & 100 & 0.3 & 0.2 & 100 & 81 & 52 & 75 & 51 & 55\\
SMC & 100 & 0.3 & 0.4 & 1 & 82 & 55 & 74 & 54 & 59\\
SMC & 100 & 0.3 & 0.4 & 100 & 82 & 55 & 72 & 52 & 58\\
SMC & 100 & 0.5 & 0.2 & 1 & 75 & 59 & 66 & 58 & 62\\
SMC & 100 & 0.5 & 0.2 & 100 & 76 & 59 & 66 & 57 & 62\\
SMC & 100 & 0.5 & 0.4 & 1 & 79 & 64 & 71 & 62 & 67\\
SMC & 100 & 0.5 & 0.4 & 100 & 79 & 64 & 71 & 62 & 67\\
 \hline 
$10^{-3}$ & 110 & 0.1 & 0.2 & 1 & 109 & 55 & 109 & 49 & 56\\
$10^{-3}$ & 110 & 0.1 & 0.2 & 100 & 109 & 56 & 109 & 52 & 59\\
$10^{-3}$ & 110 & 0.1 & 0.4 & 1 & - & - & 109 & - & 99\\
$10^{-3}$ & 110 & 0.1 & 0.4 & 100 & - & - & 109 & - & 99\\
$10^{-3}$ & 110 & 0.3 & 0.2 & 1 & 109 & 60 & 109 & 56 & 61\\
$10^{-3}$ & 110 & 0.3 & 0.2 & 100 & 109 & 60 & 109 & 50 & 50\\
$10^{-3}$ & 110 & 0.3 & 0.4 & 1 & - & - & 109 & - & 102\\
$10^{-3}$ & 110 & 0.3 & 0.4 & 100 & - & - & 109 & - & 102\\
$10^{-3}$ & 110 & 0.5 & 0.2 & 1 & 109 & 66 & 109 & 53 & 53\\
$10^{-3}$ & 110 & 0.5 & 0.2 & 100 & 109 & 68 & 109 & 64 & 69\\
$10^{-3}$ & 110 & 0.5 & 0.4 & 1 & - & - & 109 & - & 104\\
$10^{-3}$ & 110 & 0.5 & 0.4 & 100 & - & - & 109 & - & 104\\
 \hline 
$10^{-2}$ & 110 & 0.1 & 0.2 & 1 & 107 & 55 & 106 & 49 & 56\\
$10^{-2}$ & 110 & 0.1 & 0.2 & 100 & 107 & 57 & 107 & 52 & 58\\
$10^{-2}$ & 110 & 0.1 & 0.4 & 1 & 108 & 96 & 108 & - & 96\\
$10^{-2}$ & 110 & 0.1 & 0.4 & 100 & 108 & 96 & 108 & - & 96\\
$10^{-2}$ & 110 & 0.3 & 0.2 & 1 & 107 & 62 & 106 & 48 & 48\\
$10^{-2}$ & 110 & 0.3 & 0.2 & 100 & 108 & 61 & 107 & 50 & 50\\
$10^{-2}$ & 110 & 0.3 & 0.4 & 1 & 108 & 101 & 108 & - & 101\\
$10^{-2}$ & 110 & 0.3 & 0.4 & 100 & 108 & 101 & 108 & - & 101\\
$10^{-2}$ & 110 & 0.5 & 0.2 & 1 & 107 & 67 & 106 & 64 & 68\\
$10^{-2}$ & 110 & 0.5 & 0.2 & 100 & 107 & 68 & 106 & 64 & 69\\
$10^{-2}$ & 110 & 0.5 & 0.4 & 1 & 108 & 104 & 108 & - & 103\\
$10^{-2}$ & 110 & 0.5 & 0.4 & 100 & 108 & 104 & 108 & - & 103\\
 \hline 
$10^{-1}$ & 110 & 0.1 & 0.2 & 1 & 100 & 51 & 97 & 46 & 52\\
$10^{-1}$ & 110 & 0.1 & 0.2 & 100 & 100 & 53 & 96 & 48 & 54\\

\end{tabular}
\newpage
\begin{tabular}{ p{1cm}p{1cm}p{1cm}p{1cm}p{1cm}|p{1cm} p{1cm}|p{1cm}p{1cm}p{1cm}}
\hline
\hline
 $Z$ & $M_\text{i}$ & \aOv & \OmOmC & $\alpha_{\text{sc}}$ & $M_{\text{TAMS}}$ & $M_{\text{He}}$ & $M_{\text{f}}$ & $M_{\text{CO}}$ & $M_{\text{He}}$ \\ 
\hline

$10^{-1}$ & 110 & 0.1 & 0.4 & 1 & 97 & 60 & 89 & 58 & 64\\
$10^{-1}$ & 110 & 0.1 & 0.4 & 100 & 97 & 61 & 88 & 57 & 64\\
$10^{-1}$ & 110 & 0.3 & 0.2 & 1 & 97 & 60 & 93 & 58 & 63\\
$10^{-1}$ & 110 & 0.3 & 0.2 & 100 & 97 & 60 & 93 & 58 & 62\\
$10^{-1}$ & 110 & 0.3 & 0.4 & 1 & 96 & 67 & 85 & 63 & 69\\
$10^{-1}$ & 110 & 0.3 & 0.4 & 100 & 96 & 67 & 87 & 63 & 69\\
$10^{-1}$ & 110 & 0.5 & 0.2 & 1 & 94 & 67 & 89 & 65 & 70\\
$10^{-1}$ & 110 & 0.5 & 0.2 & 100 & 94 & 67 & 89 & 65 & 70\\
$10^{-1}$ & 110 & 0.5 & 0.4 & 1 & 95 & 76 & 84 & 73 & 79\\
$10^{-1}$ & 110 & 0.5 & 0.4 & 100 & 95 & 76 & 84 & 73 & 79\\
 \hline 
SMC & 110 & 0.1 & 0.2 & 1 & 95 & 49 & 90 & 46 & 51\\
SMC & 110 & 0.1 & 0.2 & 100 & 95 & 49 & 90 & 45 & 50\\
SMC & 110 & 0.1 & 0.4 & 1 & 90 & 54 & 83 & 52 & 58\\
SMC & 110 & 0.1 & 0.4 & 100 & 90 & 54 & 80 & 50 & 57\\
SMC & 110 & 0.3 & 0.2 & 1 & 86 & 58 & 79 & 57 & 62\\
SMC & 110 & 0.3 & 0.2 & 100 & 87 & 58 & 79 & 57 & 62\\
SMC & 110 & 0.3 & 0.4 & 1 & 89 & 62 & 78 & 60 & 65\\
SMC & 110 & 0.3 & 0.4 & 100 & 89 & 62 & 76 & 59 & 64\\
SMC & 110 & 0.5 & 0.2 & 1 & 83 & 66 & 73 & 64 & 69\\
SMC & 110 & 0.5 & 0.2 & 100 & 83 & 66 & 73 & 64 & 69\\
SMC & 110 & 0.5 & 0.4 & 1 & 86 & 71 & 78 & 69 & 74\\
SMC & 110 & 0.5 & 0.4 & 100 & 86 & 71 & 78 & 69 & 74\\
 \hline 
$10^{-3}$ & 120 & 0.1 & 0.2 & 1 & 119 & 60 & 119 & 54 & 60\\
$10^{-3}$ & 120 & 0.1 & 0.2 & 100 & 119 & 68 & 119 & 65 & 71\\
$10^{-3}$ & 120 & 0.1 & 0.4 & 1 & - & - & 119 & - & 103\\
$10^{-3}$ & 120 & 0.1 & 0.4 & 100 & - & - & 119 & - & 105\\
$10^{-3}$ & 120 & 0.3 & 0.2 & 1 & 119 & 65 & 119 & 58 & 64\\
$10^{-3}$ & 120 & 0.3 & 0.2 & 100 & 119 & 65 & 119 & 54 & 58\\
$10^{-3}$ & 120 & 0.3 & 0.4 & 1 & - & - & 119 & - & 111\\
$10^{-3}$ & 120 & 0.3 & 0.4 & 100 & - & - & 119 & - & 111\\
$10^{-3}$ & 120 & 0.5 & 0.2 & 1 & 119 & 75 & 119 & 71 & 77\\
$10^{-3}$ & 120 & 0.5 & 0.2 & 100 & 119 & 75 & 119 & 71 & 77\\
$10^{-3}$ & 120 & 0.5 & 0.4 & 1 & - & - & 119 & - & 114\\
$10^{-3}$ & 120 & 0.5 & 0.4 & 100 & - & - & 119 & - & 114\\
 \hline 
$10^{-2}$ & 120 & 0.1 & 0.2 & 1 & 117 & 61 & 116 & 55 & 61\\
$10^{-2}$ & 120 & 0.1 & 0.2 & 100 & 117 & 68 & 116 & 63 & 70\\
$10^{-2}$ & 120 & 0.1 & 0.4 & 1 & 118 & 105 & 118 & - & 105\\
$10^{-2}$ & 120 & 0.1 & 0.4 & 100 & 118 & 95 & 118 & - & 100\\
$10^{-2}$ & 120 & 0.3 & 0.2 & 1 & 117 & 66 & 116 & 60 & 64\\
$10^{-2}$ & 120 & 0.3 & 0.2 & 100 & 117 & 66 & 114 & 78 & 84\\
$10^{-2}$ & 120 & 0.3 & 0.4 & 1 & 118 & 110 & 118 & - & 110\\
$10^{-2}$ & 120 & 0.3 & 0.4 & 100 & 118 & 110 & 118 & - & 110\\
$10^{-2}$ & 120 & 0.5 & 0.2 & 1 & 117 & 75 & 116 & 71 & 76\\
$10^{-2}$ & 120 & 0.5 & 0.2 & 100 & 117 & 75 & 115 & 72 & 77\\
$10^{-2}$ & 120 & 0.5 & 0.4 & 1 & 118 & 113 & 118 & - & 113\\
$10^{-2}$ & 120 & 0.5 & 0.4 & 100 & 118 & 113 & 118 & - & 113\\
 \hline 
$10^{-1}$ & 120 & 0.1 & 0.2 & 1 & 108 & 56 & 105 & 51 & 57\\
$10^{-1}$ & 120 & 0.1 & 0.2 & 100 & 108 & 57 & 105 & 52 & 58\\
$10^{-1}$ & 120 & 0.1 & 0.4 & 1 & 106 & 65 & 98 & 61 & 68\\
$10^{-1}$ & 120 & 0.1 & 0.4 & 100 & 105 & 68 & 95 & 65 & 72\\
$10^{-1}$ & 120 & 0.3 & 0.2 & 1 & 105 & 66 & 101 & 64 & 69\\
$10^{-1}$ & 120 & 0.3 & 0.2 & 100 & 105 & 66 & 101 & 64 & 69\\
$10^{-1}$ & 120 & 0.3 & 0.4 & 1 & 105 & 73 & 93 & 69 & 75\\
$10^{-1}$ & 120 & 0.3 & 0.4 & 100 & 105 & 73 & 92 & 68 & 75\\

\end{tabular}
\newpage
\begin{tabular}{ p{1cm}p{1cm}p{1cm}p{1cm}p{1cm}|p{1cm} p{1cm}|p{1cm}p{1cm}p{1cm}}
\hline
\hline
 $Z$ & $M_\text{i}$ & \aOv & \OmOmC & $\alpha_{\text{sc}}$ & $M_{\text{TAMS}}$ & $M_{\text{He}}$ & $M_{\text{f}}$ & $M_{\text{CO}}$ & $M_{\text{He}}$ \\ 
\hline

$10^{-1}$ & 120 & 0.5 & 0.2 & 1 & 102 & 74 & 95 & 72 & 77\\
$10^{-1}$ & 120 & 0.5 & 0.2 & 100 & 102 & 74 & 95 & 72 & 77\\
$10^{-1}$ & 120 & 0.5 & 0.4 & 1 & 104 & 83 & 90 & 80 & 85\\
$10^{-1}$ & 120 & 0.5 & 0.4 & 100 & 104 & 83 & 90 & 80 & 85\\
 \hline 
SMC & 120 & 0.1 & 0.2 & 1 & 103 & 54 & 97 & 50 & 56\\
SMC & 120 & 0.1 & 0.2 & 100 & 103 & 55 & 97 & 50 & 55\\
SMC & 120 & 0.1 & 0.4 & 1 & 98 & 59 & 89 & 57 & 63\\
SMC & 120 & 0.1 & 0.4 & 100 & 98 & 60 & 88 & 56 & 63\\
SMC & 120 & 0.3 & 0.2 & 1 & 91 & 64 & 78 & 63 & 68\\
SMC & 120 & 0.3 & 0.2 & 100 & 91 & 64 & 78 & 63 & 68\\
SMC & 120 & 0.3 & 0.4 & 1 & 97 & 68 & 83 & 66 & 72\\
SMC & 120 & 0.3 & 0.4 & 100 & 97 & 68 & 83 & 65 & 71\\
SMC & 120 & 0.5 & 0.2 & 1 & 91 & 73 & 80 & 70 & 76\\
SMC & 120 & 0.5 & 0.2 & 100 & 91 & 73 & 80 & 71 & 76\\
SMC & 120 & 0.5 & 0.4 & 1 & 92 & 78 & 86 & 76 & 81\\
SMC & 120 & 0.5 & 0.4 & 100 & 92 & 78 & 86 & 76 & 81\\
 \hline 
$10^{-3}$ & 150 & 0.1 & 0.2 & 1 & 149 & 79 & 148 & 70 & 79\\
$10^{-3}$ & 150 & 0.1 & 0.2 & 100 & 149 & 84 & 148 & 77 & 85\\
$10^{-3}$ & 150 & 0.1 & 0.4 & 1 & - & - & 149 & - & 130\\
$10^{-3}$ & 150 & 0.1 & 0.4 & 100 & - & - & 149 & - & 129\\
$10^{-3}$ & 150 & 0.3 & 0.2 & 1 & 149 & 86 & 149 & 64 & 64\\
$10^{-3}$ & 150 & 0.3 & 0.2 & 100 & 149 & 84 & 149 & 72 & 80\\
$10^{-3}$ & 150 & 0.3 & 0.4 & 1 & - & - & 149 & - & 129\\
$10^{-3}$ & 150 & 0.3 & 0.4 & 100 & - & - & 149 & - & 130\\
$10^{-3}$ & 150 & 0.5 & 0.2 & 1 & 149 & 97 & 148 & 92 & 98\\
$10^{-3}$ & 150 & 0.5 & 0.2 & 100 & 149 & 97 & 148 & 92 & 98\\
$10^{-3}$ & 150 & 0.5 & 0.4 & 1 & - & - & 149 & - & 140\\
$10^{-3}$ & 150 & 0.5 & 0.4 & 100 & - & - & 149 & - & 140\\
 \hline 
$10^{-2}$ & 150 & 0.1 & 0.2 & 1 & 146 & 79 & 145 & 71 & 79\\
$10^{-2}$ & 150 & 0.1 & 0.2 & 100 & 146 & 80 & 145 & 72 & 80\\
$10^{-2}$ & 150 & 0.1 & 0.4 & 1 & 147 & 123 & 147 & - & 123\\
$10^{-2}$ & 150 & 0.1 & 0.4 & 100 & 147 & 129 & 147 & - & 129\\
$10^{-2}$ & 150 & 0.3 & 0.2 & 1 & 146 & 84 & 145 & 69 & 69\\
$10^{-2}$ & 150 & 0.3 & 0.2 & 100 & 146 & 83 & 142 & 100 & 107\\
$10^{-2}$ & 150 & 0.3 & 0.4 & 1 & 147 & 131 & 147 & - & 131\\
$10^{-2}$ & 150 & 0.3 & 0.4 & 100 & 147 & 121 & 147 & - & 122\\
$10^{-2}$ & 150 & 0.5 & 0.2 & 1 & 145 & 96 & 144 & 91 & 98\\
$10^{-2}$ & 150 & 0.5 & 0.2 & 100 & 146 & 96 & 144 & 91 & 98\\
$10^{-2}$ & 150 & 0.5 & 0.4 & 1 & - & - & 147 & - & 134\\
$10^{-2}$ & 150 & 0.5 & 0.4 & 100 & 147 & 134 & 147 & - & 134\\
 \hline 
$10^{-1}$ & 150 & 0.1 & 0.2 & 1 & 134 & 73 & 129 & 67 & 74\\
$10^{-1}$ & 150 & 0.1 & 0.2 & 100 & 134 & 75 & 129 & 68 & 75\\
$10^{-1}$ & 150 & 0.1 & 0.4 & 1 & 131 & 84 & 119 & 79 & 87\\
$10^{-1}$ & 150 & 0.1 & 0.4 & 100 & 131 & 84 & 116 & 78 & 86\\
$10^{-1}$ & 150 & 0.3 & 0.2 & 1 & 129 & 85 & 123 & 82 & 88\\
$10^{-1}$ & 150 & 0.3 & 0.2 & 100 & 129 & 85 & 123 & 82 & 88\\
$10^{-1}$ & 150 & 0.3 & 0.4 & 1 & 130 & 91 & 114 & 85 & 92\\
$10^{-1}$ & 150 & 0.3 & 0.4 & 100 & 130 & 91 & 114 & 84 & 93\\
$10^{-1}$ & 150 & 0.5 & 0.2 & 1 & 118 & 95 & 104 & 92 & 98\\
$10^{-1}$ & 150 & 0.5 & 0.2 & 100 & 118 & 95 & 103 & 92 & 98\\
$10^{-1}$ & 150 & 0.5 & 0.4 & 1 & 128 & 104 & 112 & 100 & 107\\
$10^{-1}$ & 150 & 0.5 & 0.4 & 100 & 128 & 104 & 112 & 100 & 106\\
 \hline 

\end{tabular}
\newpage
\begin{tabular}{ p{1cm}p{1cm}p{1cm}p{1cm}p{1cm}|p{1cm} p{1cm}|p{1cm}p{1cm}p{1cm}}
\hline
\hline
 $Z$ & $M_\text{i}$ & \aOv & \OmOmC & $\alpha_{\text{sc}}$ & $M_{\text{TAMS}}$ & $M_{\text{He}}$ & $M_{\text{f}}$ & $M_{\text{CO}}$ & $M_{\text{He}}$ \\ 
\hline

SMC & 150 & 0.1 & 0.2 & 1 & 124 & 71 & 116 & 67 & 74\\
SMC & 150 & 0.1 & 0.2 & 100 & 122 & 72 & 113 & 66 & 73\\
SMC & 150 & 0.1 & 0.4 & 1 & 120 & 77 & 105 & 74 & 81\\
SMC & 150 & 0.1 & 0.4 & 100 & 120 & 77 & 99 & 72 & 80\\
SMC & 150 & 0.3 & 0.2 & 1 & 110 & 82 & 91 & 81 & 87\\
SMC & 150 & 0.3 & 0.2 & 100 & 110 & 82 & 93 & 81 & 87\\
SMC & 150 & 0.3 & 0.4 & 1 & 120 & 88 & 97 & 86 & 92\\
SMC & 150 & 0.3 & 0.4 & 100 & 119 & 88 & 100 & 85 & 92\\
SMC & 150 & 0.5 & 0.2 & 1 & 115 & 93 & 101 & 90 & 97\\
SMC & 150 & 0.5 & 0.2 & 100 & 115 & 94 & 101 & 90 & 97\\
SMC & 150 & 0.5 & 0.4 & 1 & 112 & 98 & 107 & 95 & 101\\
SMC & 150 & 0.5 & 0.4 & 100 & 112 & 98 & 107 & 95 & 101\\
 \hline 
 \hline
\end{tabular}

\twocolumn

\section{Resolution Test}
\label{Ap-ResTest}

\begin{figure*}
   \centerline{\hspace*{0.015\textwidth}
               \includegraphics[width=0.40\textwidth,clip=]{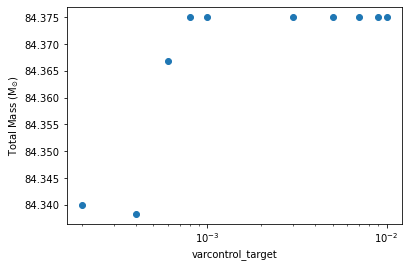}
               \hspace*{0.03\textwidth}
               \includegraphics[width=0.40\textwidth,clip=]{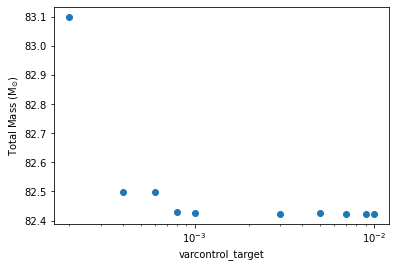}
              }
     \vspace{-0.35\textwidth}   
     \centerline{\Large       
         \hfill}
     \vspace{0.31\textwidth}
\caption{Final mass of stellar models run in parallel for the same stage of evolution. The left panel is for main sequence evolution, while the right panel is for TAMS until core-Helium exhaustion. All models were started from the same initial model, except for a different resolution parameter - in this case, \textit{varcontrol\_target}.}
\label{Fig-FinalMassVarcon}
\end{figure*}

In order to ensure that our chosen resolution was converged, we conducted a resolution test which we define here. As of version r15140, MESA has two forms of top-level resolution definition for controlling temporal and spatial resolutions - these are \textit{time\_delta\_coefficient} and \textit{mesh\_delta\_coefficient}, though there are many more parameters which exist (for examples and included resolution tests, there is \cite{Farmer19} and \cite{Costa22}). This version is the first to include the \textit{time\_delta\_coefficient} parameter. Previously, resolution control for the time domain was achieved through varying \textit{varcontrol\_target}. This is now discouraged, and a limit of \textit{varcontrol\_target $> 1d-4$} is imposed by default, though this can be overridden. For completeness, variations in \textit{varcontrol\_target} were also tested. 

\begin{figure*}
   \centerline{\hspace*{0.015\textwidth}
               \includegraphics[width=0.40\textwidth,clip=]{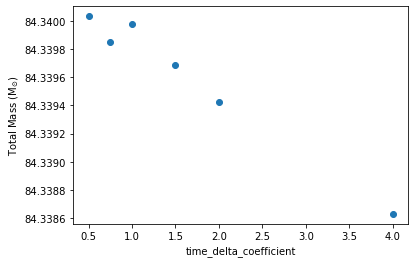}
               \hspace*{0.03\textwidth}
               \includegraphics[width=0.40\textwidth,clip=]{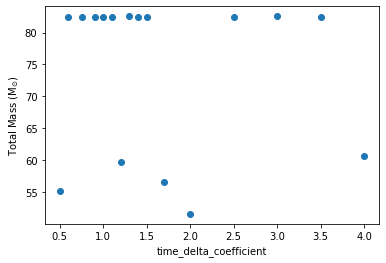}
              }
     \vspace{-0.35\textwidth}   
     \centerline{\Large       
         \hfill}
     \vspace{0.31\textwidth}
\caption{Similar to the above panel, except that instead of \textit{varcontrol\_target}, the varying resolution parameter is \textit{time\_delta\_coeff}. The bifurcation of results can be seen in the right panel, covering TAMS to core-Helium exhaustion. Where some results would remain above $80 \ M_{\odot}$, others would lose as much as $30 \ M_{\odot}$ of stellar material.}
\label{Fig-FinalMassTDC}
\end{figure*}

\begin{figure*}
   \centerline{\hspace*{0.015\textwidth}
               \includegraphics[width=0.40\textwidth,clip=]{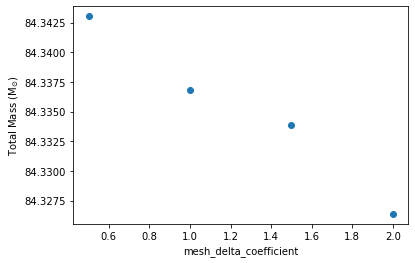}
               \hspace*{0.03\textwidth}
               \includegraphics[width=0.40\textwidth,clip=]{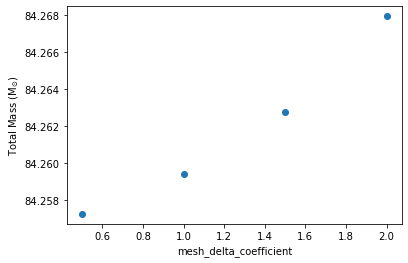}
              }
     \vspace{-0.35\textwidth}   
     \centerline{\Large       
         \hfill}
     \vspace{0.31\textwidth}
\caption{Initial tests conducted with varying \textit{mesh\_delta\_coeff} for each stage. Hydrogen burning is shown on the left, while Helium burning is on the right as before.}
\label{Fig-FinalMassMDC}
\end{figure*}

To attempt to optimise the resolution which would eventually be chosen, the evolution of the star up to the end of core Helium burning was split into two stages. In this way, the resolution could be tailored to each burning stage to provide converged results as quickly as possible. The first stage covered the star's evolution from Hydrogen ignition at the Zero Age Main Sequence (ZAMS) to the Terminal Age Main Sequence (TAMS). Picking up where the first stage left off, the second stage continues the life of the star to the end of Helium burning in the core.

When splitting the model, we took advantage of a feature present in MESA which allows a model to be saved upon termination of the run. However, this does not save the internal profile of the model upon termination, unlike the \textit{photo} files. MESA warns that this is the case, and further suggests that this shouldn't cause particularly egregious divergences between a model run from the \textit{.mod} file or run as one continuous model. However, for our test case, it was found that considerable differences could be seen in the later stages of evolution for stars run in our grid.
For the purposes of the resolution test, since the tests were to examine convergence, using the \textit{.mod} files was acceptable as the physics does not change, merely the internal structure.

Three series of tests were conducted. One which varied \textit{time\_delta\_coefficient}, another varying \textit{varcontrol\_target} and a final smaller series varying \textit{mesh\_delta\_coefficient}. During the Main Sequence stage, all resolutions showed consistent results for the final properties of the model, which can be seen in the left-hand panels of Figures \ref{Fig-FinalMassVarcon}, \ref{Fig-FinalMassTDC}, and \ref{Fig-FinalMassMDC}. While \textit{varcontrol\_target} shows a convergence towards $1d - 4$, it is noted that the range of values which this encompasses are small in comparison to the total model. The spatial resolution test, \textit{mesh\_delta\_coefficient}, showed little variation between models as well at this stage, as did \textit{time\_delta\_coefficient}. 

When moving into the core Helium burning stage, \textit{time\_delta\_coeff} showed a bifurcation in the results, showcased in the right-hand panel of Figure \ref{Fig-FinalMassTDC}. While most models would act typically, and complete core Helium burning in an expected manner (herein referred to as ``Case 1"), some models would appear to continue past Helium burning. For the models which kept going (herein described as ``Case 2"), the error started near the end of where Helium burning should've ended, at which point a small error regarding minimum abundances propagated. These errors compounded, which eventually leads to the model continuing past core Helium burning into core Carbon burning. If running the MESA ``Vink" wind recipe, the model will lose as much as $30 \ M_{\odot}$. However, if using the corrected iron abundance recipe mentioned before, the star will still lose mass, but only $2 \sim 3 \ M_{\odot}$ in comparison. This is due to how each recipe handles the metallicity dependence of \cite{Vink01}. In default MESA ''Vink" wind, this recipe accounts for all metals present at the surface of the star when calculating the mass loss in each timestep. This however, is incorrect, and instead the recipe should only depend on surface iron abundances.

Upon deeper investigation, this issue was caused by use of the ``exponential" overshooting prescription in MESA. Thus, by using the ``step" overshooting prescription in MESA, the models would not encounter the same issues with regards to abundance changes, and the follow-on errors which occurred. Exponential overshooting will produce a gradient of elements and products of the CNO cycle above the core as this is mixed in gradually. If the shell dips downwards in to the core during the early core Helium burning stage, then this would cause a change of one of the CNO elements in the Hydrogen-burning shell region's boundary. As the shell region is fully convective, and full convection results in instantaneous mixing. This leads to an overabundance of a CNO element throughout the shell. MESA would then correct for this by increasing the abundances of the other elements - over-correcting the abundance. When this occurs, the model evolves as per the Case 2 model, rather than the typical Case 1 model. Some time after, the entire star becomes convective, which mixes the high abundance of metals all the way to the surface, where the wind is calculated, which thus produces the unrealistically high mass-loss rates.

Whereas exponential overshooting, by its name, exponentially reduces the overshooting over a distance parameter $f_0$, step overshooting is instead a simple function of extending the core by a distance described by $\alpha_{\text{ov}}$. This means that all elements in this region are fully mixed, and so when the shell touches the step overshooting region as it does in the Case 2 models mentioned before, there is no gradient of elements that the shell then corrects for. This is then fully mixed, which avoids the correction.

Given the results of our resolution test, we decide to use a value of \textit{varcontrol\_target} of $2d-4$, while both \textit{time\_delta\_coefficient} and \textit{mesh\_delta\_coefficient} were left at their default values. Additionally, we use the step overshooting prescription.

\section{High $\Omega$ models, and $\Omega$ at low $Z$}
\label{Ap-RotationDiscussion}

 {In Section \ref{s-ParamSpace}, we} have discussed the effects we have observed in our models, including the effects of low versus high rotation. Another effect which we have also discussed is the spin-up to supercritical breakup speeds for rotation as a function of decreasing metallicity, the conclusion of which is illustrated by Figure \ref{Fig-Rot_Z_sketch}. With this appendix, we will summarise our results in the context of current and potential future work. 

As mentioned, we find a metallicity-dependent limit of rotation which could prohibit formation of {heavy} black holes at lower metallicity. We also expect the distribution of rotation rates to shift upwards at lower metallicity due to the lack of angular momentum loss through winds, which would further truncate the limit on the number of black holes with $M_{\text{BH}} > 50$ $M_{\odot}$. 

There is evidence from earlier works such as \cite{Meynet06} which show that a star which reaches breakup velocity and become unbound experiences a period of intense mass loss until the remaining star can become gravitationally bound again. For models in \cite{Meynet06}, this was achieved by removing supercritical layers, then allowing the star to stabilise, which resulted in mass loss in the range of $35 \sim 42$ $M_{\odot}$ from a $60$ $M_{\odot}$ star at varying metallicity and rotation rates (see Table 1 from aforementioned paper). We can estimate from this that the mass loss for our stars, which are the same mass or greater than those of \cite{Meynet06} would be similarly dramatic.

In a small subset of models, we implemented the $\Gamma \Omega$ mass loss boost factor of \cite{Maeder00B} which we find can dramatically change the effect of rotation in the high rotation, low metallicity regime. Firstly, the effect of this boost allows the model to avoid becoming supercritical by boosting the mass loss as a factor {given in Equation \ref{Eq-MM2000B_Mdot}}, which increases dramatically as the rotation rate gets close to critical, thus allowing more angular momentum to be lost before the model becomes supercritical. This was seen to stop models from reaching breakup speeds up to initial rotation values of \OmOmC $= 0.8$. 

The MESA implementation of rotation boosted mass loss is from \cite{Heger00} and \cite{Langer98}, where mass loss is boosted by a parameter of $[1 / (1 - $ \OmOmC $)] ^{\xi}$, where $\xi$ is the {fit value}, nominally $\xi = 0.43$. This, however, does not account for the von Zeipel theorem \citep{Maeder00B}. In \cite{Maeder00B}, the equation of rotation boosted mass loss is as follows;

\begin{equation}
    \label{Eq-MM2000B_Mdot}
    \frac{ \dot{M}(\Omega) }{ \dot{M}(0) } \simeq \frac{ ( 1 - \Gamma )^{\frac{1}{\alpha} -1} }{ \left[ 1 - \frac{\Omega^2}{2 \pi G \rho_{\text{m}}} - \Gamma \right]^{\frac{1}{\alpha} - 1} }
\end{equation}

which is given in their equation 4.29 and correctly accounts for the von Zeipel theorem.

Rotation in MESA is handled as a diffusive mixing process (\cite{Paxton13}, Section 6.1.2 for MESA implementation, and references therein for complete physical descriptions) compared to other stellar evolution codes which handle rotation as diffusion-advection approach, notably the GENEC stellar evolution code \citep{Geneva08}. The main difference between these two methods arises in the transport of angular momentum, as the diffusive process determined used by MESA considers various instabilities, discussed briefly in Section \ref{s-Methods} of this paper, and in more detail in \cite{Heger00}, while GENEC uses the formulae of \cite{Maeder98}.

Overall, we cannot say exactly where the maximum \OmOmC limit is in the context of wider physical processes - in our grid, this appears at \OmOmC $= 0.4$ and decreases with metallicity. Results from previous work suggests this limiting value could be higher, while {a different} stellar evolution code could suggest a different number. Thus, at this time, we have provided a fiducial investigation which informs future work.

\section{Enhanced Winds in close proximity to the Eddington Limit}
\label{Ap-Winds}

\begin{figure}
  \centering
  \includegraphics[width=1\linewidth]{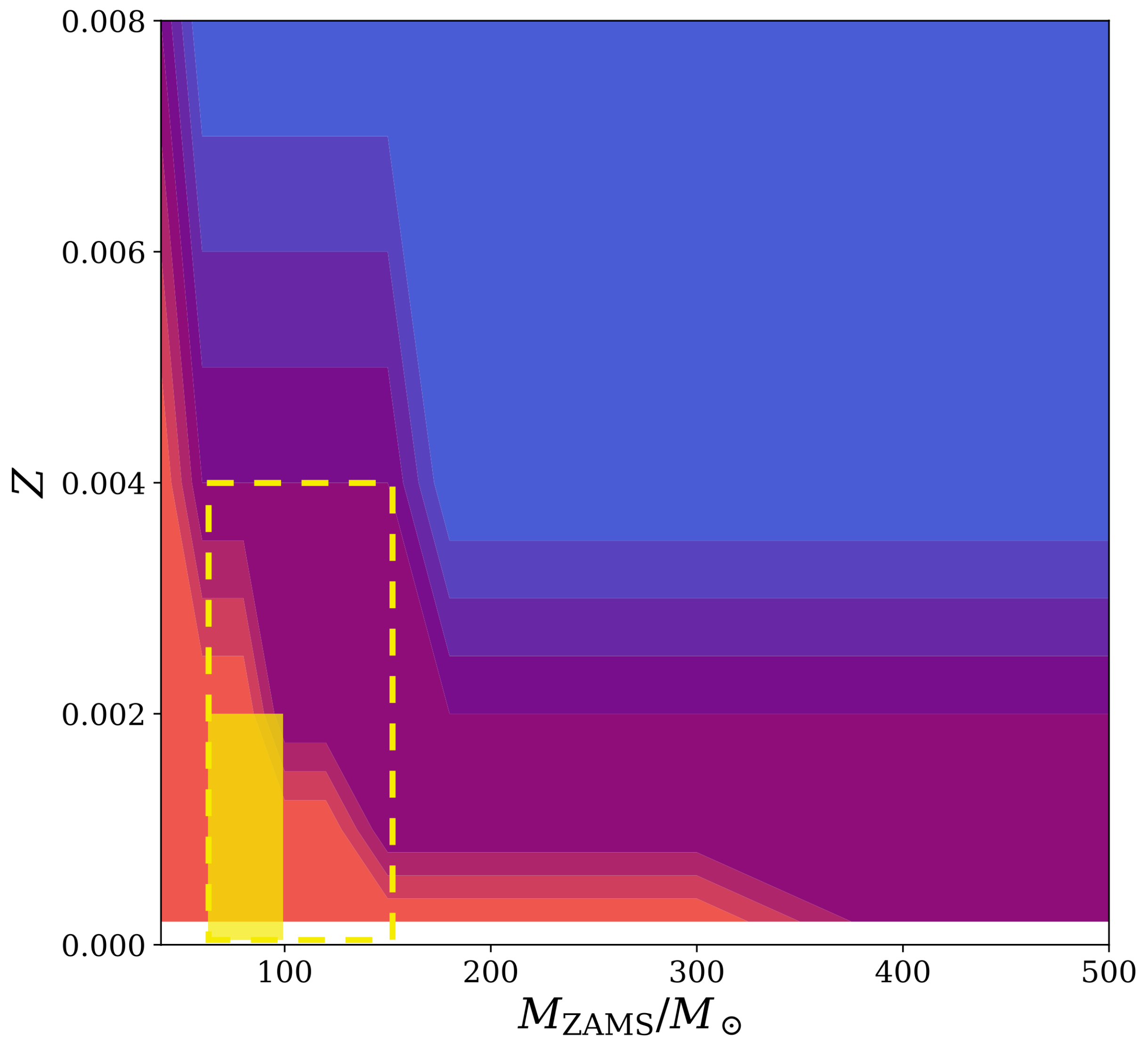}
\caption{Sketch of optical wind thickness as a factor of metallicity and ZAMS mass. The blue region denotes the region where the star has a high $\Gamma_{\rm e}$ for the entire main sequence, while the orange region describes a star has a low $\Gamma_{\rm e}$ for the entire main sequence. Purple has a high $\Gamma_{\rm e}$ on the cool side of the bi-stability jump \citep{Sab23}. Our grid is bounded by the yellow dashed line, while the models with highest final masses ($M_{\text{final}} > 50 M_\odot$) and $M_\text{CO} < 36.3 M_\odot$ are denoted by the yellow box.}
\label{Fig-GauthamPlot}
\end{figure}

{As discussed in Sect.\,\ref{s-Methods}, the mass-loss recipe employed in this study is identical to the one in \cite{Vink21}. Since that work, we have implemented mass-loss rate relations for very massive stars in close proximity to the Eddington limit from \cite{Vink11} for both high and low $Z$ \citep{Sabhahit22,Sab23}. 

Figure\,\ref{Fig-GauthamPlot} indicates the parameter space of these very massive star models for which a new mass-loss recipe was implemented, in comparison to the parameter space of the current study.
The figure shows the optical thickness of winds as a function of metallicity and initial mass. 
The evolutionary models from \cite{Vink21} that produce the highest final masses and BH masses are highlighted by the yellow box. These winds remain optically thin and a mass-loss boost would not be applied. 

The entire parameter space of the current models is 
encapsulated by the yellow dashed line, where the winds are either totally or partially optically thin (as shown in the orange background). These models are on the verge of becoming optically thick. Comparison of the models in our current study compared to those of \citep{Sabhahit22,Sab23} shows that our results are not affected by a mass-loss boost.

In short, the reason why the mass-loss recipe applied in the current study and the ones in \cite{Sabhahit22,Sab23} are equivalent is that the parameter space of our current study is almost entirely in the optically thin regime of \cite{Sabhahit22,Sab23}.}

\section{Inflation and MLT++ at low metallicity}
\label{Ap-MLT++_low_Z}

MLT++ is known to affect the inflation characteristics of MESA models, however it is also necessary to allow massive star models to evolve without convergence issues (as discussed in Section \ref{ss-MESAMod}). We ran a \Mi $= 93$ $M_{\odot}$ model with low overshooting ($\sim 0$) and without MLT++ in order to test if the maximum BH mass we predict is robust.
The reason this is an appropriate test is that at very low metallicity you would not expect to witness a large envelope inflation. However, at a higher metallicity ($1/50$ $Z_{\odot}$), massive stars with $M_{\rm ZAMS}$ above $80$ $M_{\odot}$ may evolve into supergiants during core Hydrogen burning \citep{2015A&A...581A..15S}. 
We apply a very low metallicity ($Z = 1/1000 \text{ th } Z_\odot$) from our prediction in Section \ref{s-Mbh_Eqn}. With these conditions, the model is able to remain compact for the He burning lifetime and not undergo inflation, which will not necessarily be true for the rest of the parameter space.
While the envelope mixing choice may, similarly to mass-loss assumptions as discussed in \citet{Vink21}, affect the cut-off point on the maximum BH mass with $Z$, the sheer maximum BH mass value that is expected to be achieved at the lowest $Z$ values, is not expected to depend on our decision to use MLT++.


\bsp	
\label{lastpage}
\end{document}